\documentclass[openacc]{rsproca_new}

\usepackage{graphicx}
\usepackage{color}
\usepackage[normalem]{ulem}

\newcommand{\unit}[1]{\,\mathrm{#1}}
\graphicspath{Figures}

\begin{document}

\title{Taylor-Couette flow for astrophysical purposes}

\author{
H. Ji$^{1,2}$ and J. Goodman$^{1}$}

\address{$^{1}$Department of Astrophysical Sciences, Princeton University, Princeton, New Jersey, 08544, U.S.A.\\
$^{2}$Princeton Plasma Physics Laboratory, Princeton University, Princeton, New Jersey, 08543, U.S.A.\\
}

\subject{xxxxx, xxxxx, xxxx}


\corres{Hantao Ji\\
\email{hji@pppl.gov}}

\begin{abstract}
A concise review is given of astrophysically motivated experimental and theoretical research on Taylor-Couette flow.
The flows of interest rotate differentially with inner cylinder faster than outer one but are linearly stable against Rayleigh's inviscid centrifugal instability.
At shear Reynolds numbers as large as $10^6$, hydrodynamic flows of this type (quasi-keplerian) appear to be nonlinearly stable: no turbulence is seen that cannot be attributed to interaction with the axial boundaries, rather than the radial shear itself. 
Direct numerical simulations agree, although they cannot yet reach such high Reynolds numbers.
This result indicates that accretion-disc turbulence is not purely hydrodynamic in origin, at least insofar as it is driven by radial shear.
Theory, however, predicts linear magnetohydrodynamic (MHD) instabilities in astrophysical discs: in particular, the standard magnetorotational instability (SMRI).
MHD Taylor-Couette experiments aimed at SMRI are challenged by the low magnetic Prandtl numbers of liquid metals.
High fluid Reynolds numbers and careful control of the axial boundaries are required.
The quest for laboratory SMRI has been rewarded with the discovery of some interesting inductionless cousins of SMRI, and the recently reported success in demonstrating SMRI itself by taking advantage of conducting axial boundaries.  Some outstanding questions and near-future prospects are discussed, especially in connection with astrophysics.

\end{abstract}


\maketitle

\section{Introduction}


Astrophysical interest in Taylor-Couette (hereafter ``TC'') flow is motivated by astronomical systems that rotate differentially, usually at extremely large Reynolds numbers. Stars rotate because they form by gravitational contraction from interstellar gas through some 20 orders of magnitude in density, so that any initial vorticity was greatly magnified. Indeed, stars could not form without shedding most of their initial angular momentum, by processes not fully understood \cite{mestel+spitzer56,zhao+20}. While stars are supported against their own gravity mainly by pressure and so are rather round, there is a broad category of very flattened systems called ``discs'' whose support is mainly centrifugal. Spiral galaxies belong here, as do the protoplanetary discs recently imaged by the Atacama Large Millimeter Array \cite{andrews+2018}. Discs around compact objects (white dwarfs, neutron stars, and black holes) are usually too compact to be imaged; their rotational flattening is inferred indirectly. 

The mass of the disc is usually small compared to that of its central object, leading to a keplerian angular velocity profile $\Omega\propto r^{-3/2}$, $r$ being the distance from the rotation axis (cylindrical radius).
Galactic masses are more extended, so that $\Omega\propto r^{-1}$.
For such rotation laws, the specific angular momentum $r^2\Omega$ increases outward, so that Rayleigh's centrifugal instability does not operate. For this reason, TC flows whose outer cylinder is at rest (or counter-rotates) are of limited astrophysical interest. We use the term \emph{quasi-keplerian} for all rotation laws in which $\partial\Omega/\partial r$ and $\partial(r^2\Omega)/\partial r$ have opposite signs.
For an ideal-Couette fluid rotation profile $\Omega(r)=a+br^{-2}$, this requires $ab>0$.

Most discs accrete: their inner parts flow gradually inward to join the central object.
The inflowing gas must shed angular momentum, transferring it outward through the disc via turbulent torques, or expelling angular momentum in a magnetized wind \cite{frank2002accretion}.
Absent centrifugal instability, proposed causes of disc turbulence fall into two main categories: (1) a nonlinear or subcritical hydrodynamic shear-flow instability~\cite{zeldovich81,richard99}; and (2) a linear or supercritical instability of magnetized flow known as the magnetorotational instability (MRI)~\cite{balbus91}. It turns out that both of these candidate mechanisms can be actually studied experimentally in TC flow~\cite{richard01,ji01}. 

\begin{figure}[ht]
\centering\includegraphics[width=3in]{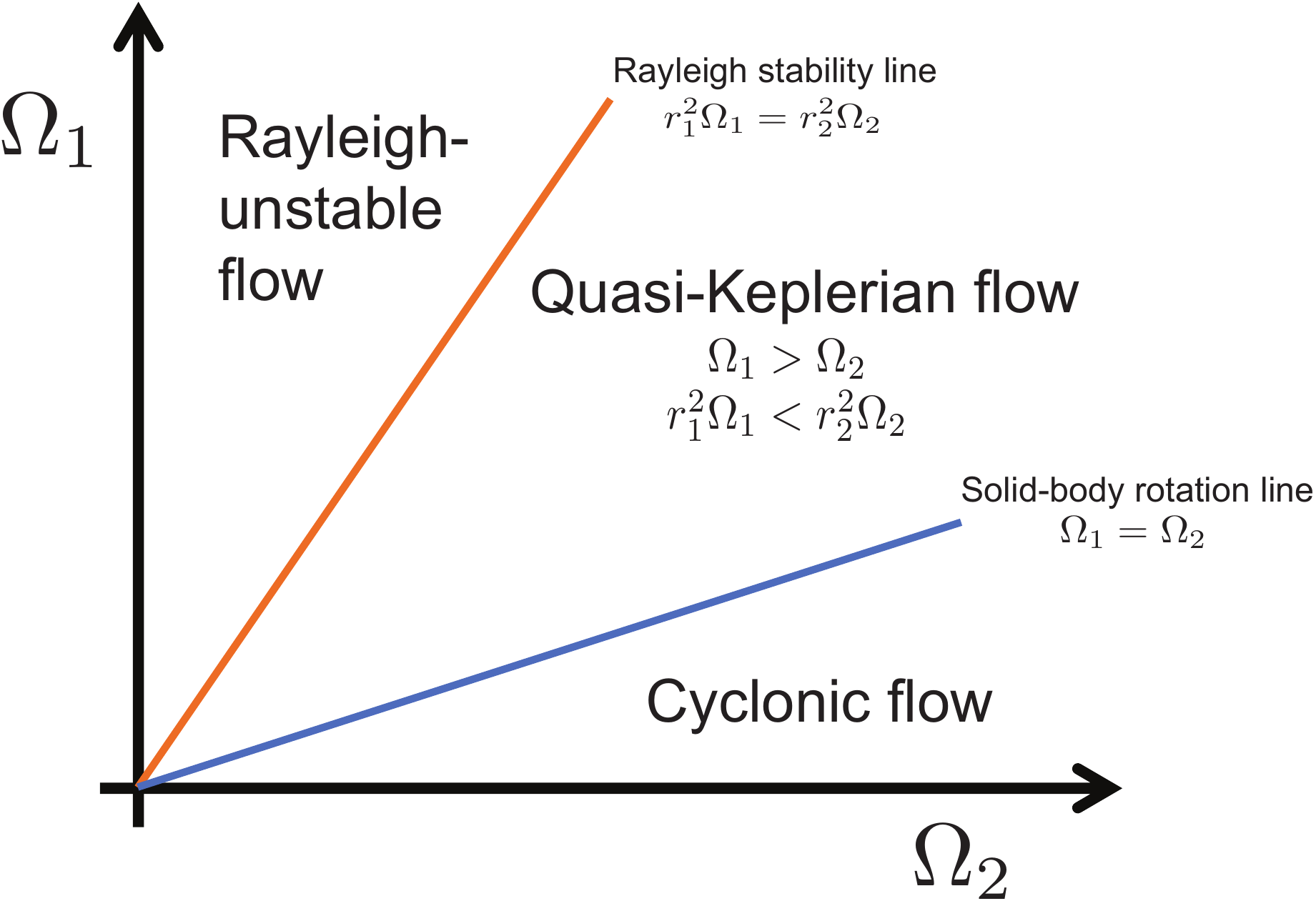}
\caption{Different Taylor-Couette (TC) flows in the plane $(\Omega_1,\Omega_2)$ of the rotation rates of the inner (1) and outer (2) cylinders. $(r_1,r_2)$ are the cylinders' radii.}
\label{fig_diagram}
\end{figure}

Viscously driven quasi-keplerian TC flow requires that both cylinders rotate.
Figure \ref{fig_diagram} illustrates the flow regimes defined by the inner ($\Omega_1$) and outer ($\Omega_2$) rotation rates. 
Both quasi-keplerian and cyclonic flows, separated by the solid-body line $\Omega_1=\Omega_2$, are centrifugally stable. 
Cyclonic flows with the inner cylinder at rest were originally used in the late $19^{\rm th}$ century by Mallock~\cite{mallock1889,mallock1896} and Couette~\cite{couette1890} to measure the viscosity of water. 
This was followed by Wendt~\cite{wendt33} and Taylor~\cite{taylor36} in 1930s, and by others in later decades. 
Other than recently predicted linear instabilities in hydrodynamics~\cite{deguchi17} and in MHD~\cite{mamatsashvili19}, cyclonic flows are subject to subcritical transition to turbulence when the shear Reynolds number is sufficiently large \cite{wendt33,taylor36,tillmark96,lesur05,feldmann22}. 
Cyclonic flow is rare in astrophysics, excepting star-disc boundaries \cite{Belyaev+Rafikov2012,Coleman+2022}. 

Centrifugally unstable and quasi-keplerian flows are separated by the Rayleigh stability line~\cite{rayleigh1916} $r_1^2\Omega_1=r_2^2\Omega_2$.
The former, which were investigated first by Mallock~\cite{mallock1889,mallock1896} and then by Taylor~\cite{taylor1923} in 1923 in great detail, are the subject of most papers in this special issue and will not be covered here. 
Quasi-keplerian flows are the main subject of this concise review.

We restrict discussion to large Reynolds numbers, defined by $Re\equiv r_1^2\Omega_1/\nu\gg 1$.
For narrow gaps ($r_2-r_1\ll r_1$), a better definition might be one explicitly based on shear.
For example, with the ideal-Couette profile, we might prefer
$Re'\equiv b/\nu=\nu^{-1}(\Omega_1-\Omega_2)/(r_1^{-2}-r_2^{-2})$. 
This is nearly the same as $Re$ when $r_1/r_2$ and $\Omega_2/\Omega_1$ are significantly less than unity.

The rest of this mini-review will first focus on hydrodynamic experiments in water, and then on magnetohydrodynamics using liquid metals. Astrophysical implications of these experiments will be discussed including their future prospects. We conclude with brief citing of TC experiments with stratification, as well as those using gas or plasma.

\section{Hydrodynamic Experiments}

\begin{table}[ht]
\caption{Experiments and modeling of  hydrodynamic TC flow in quasi-keplerian regimes.}
\label{hydro_experiments}
\begin{tabular}{llp{7.1cm}p{1.2cm}}
\hline
\bf Year & \bf Author(s) & \bf Main results in the quasi-keplerian regime & \bf Method \\
\hline\\
1933 & Wendt~\cite{wendt33} & Limited measurements of velocity profile and torque. No turbulence reported. & Exp \\
2001 & Richard~\cite{richard01} & Measured velocity profile affected by endcaps; turbulence identified by flow visualization. & Exp\\
2002 & Beckley~\cite{beckley02} & Measured torque consistent with theoretical predictions for laminar Ekman circulation & Exp\\
2004 & Kageyama et al.~\cite{kageyama04} & Measured velocity profile and spin-down time consistent with laminar Ekman circulation & Exp and modeling\\
2006 & Ji et al.~\cite{ji06,schartman12} & No signs of turbulence at $Re=2\times 10^6$ with differentially rotating segmented endcaps & Exp\\
2008 & Obabko et al.~\cite{obabko08} & Confirmed reduction of Ekman cells via segmented endcaps. & Modeling\\
2011 & Paoletti et al.~\cite{paoletti11,paoletti12} & Measured enhancement of torque over laminar flow; endcaps attached to outer cylinder & Exp\\
2012 & Avila~\cite{avila12} & Up to $Re \approx 6.4\times 10^3$; bulk flow turbulence driven by segmented endcap rings & Modeling\\
2014 & Edlund \& Ji~\cite{edlund14} & No turbulence despite active perturbations; one endcap ring and two radial rims & Exp\\
2015 & Nordsiek et al.~\cite{nordsiek15} & Enhanced torque due to axial transport of angular momentum to endcaps & Exp\\
2015 & Edlund \& Ji~\cite{edlund15} & Confirmed $Re$ self-similarity of the quiescent flow when Ekman circulation is minimized & Exp \\
2017 & Lopez \& Avila~\cite{lopez17} & Up to $Re \approx 5\times 10^4$; turbulence confined near endcaps, bulk flow laminar & Modeling\\
\hline
\end{tabular}
\vspace*{-4pt}
\end{table}

The first connection between laboratory TC and accretion-disc flows was made by Zeldovich~\cite{zeldovich81} in 1981 and was followed by Richard and Zahn~\cite{richard99} in 1999, but both were based on subcritical transition to turbulence observed in the cyclonic flows~\cite{couette1890,wendt33,taylor36}. The idea is that, keplerian flows with sufficiently large Reynolds numbers (as in astrophysics) should always be turbulent due to subcritical transitions despite linear stability, as in cyclonic flows and many other flows. In fact, to our knowledge there have been no exceptions to this hypothesis other than quasi-keplerian flows (see below).

A contentious debate broke out as to quasi-keplerian flows are indeed nonlinearly or subcritically unstable to finite-amplitude perturbations, and if so, how efficiently this mechanism transports angular momentum radially outward.
Following the initial suggestions~\cite{zeldovich81,richard99}, many efforts were made and published to study this possibility.
None were conclusive, but many authors tended to favor the idea of nonlinear instability, although there were some skeptical astrophysical theorists~\cite{balbus96,hawley99,lesur05}. Testing this idea in actual TC experiments in the quasi-keplerian regime therefore became a critical next step, which was performed by several groups in the following years and sometimes accompanied by relevant direct numerical simulations. All relevant experimental and numerical work to date is summarized in Table \ref{hydro_experiments}.

In the early years of TC flow research, the only documented experiments in the quasi-keplerian regime were performed by Wendt~\cite{wendt33}, in the final stages of a scan over $\Omega_1/\Omega_2$ starting from the centrifugally unstable regime.
The torque data were re-plotted by Coles~\cite{coles65} in his Fig.~1, but with no signs of turbulence reported.
After a 68-year hiatus, in 2001, the first research devoted specifically to TC flows in quasi-keplerian regime was published in the Ph.D. thesis of D. Richard~\cite{richard01}. 
Richard observed quasi-keplerian turbulence via Kalliroscope particles and laser-doppler velocimetry, but he did not identify hysteresis in the transitions between laminar and turbulent states, such as he himself confirmed for nonlinear instability in cyclonic flows.

\begin{figure}[ht]
{\centering
\includegraphics[width=1.4in]{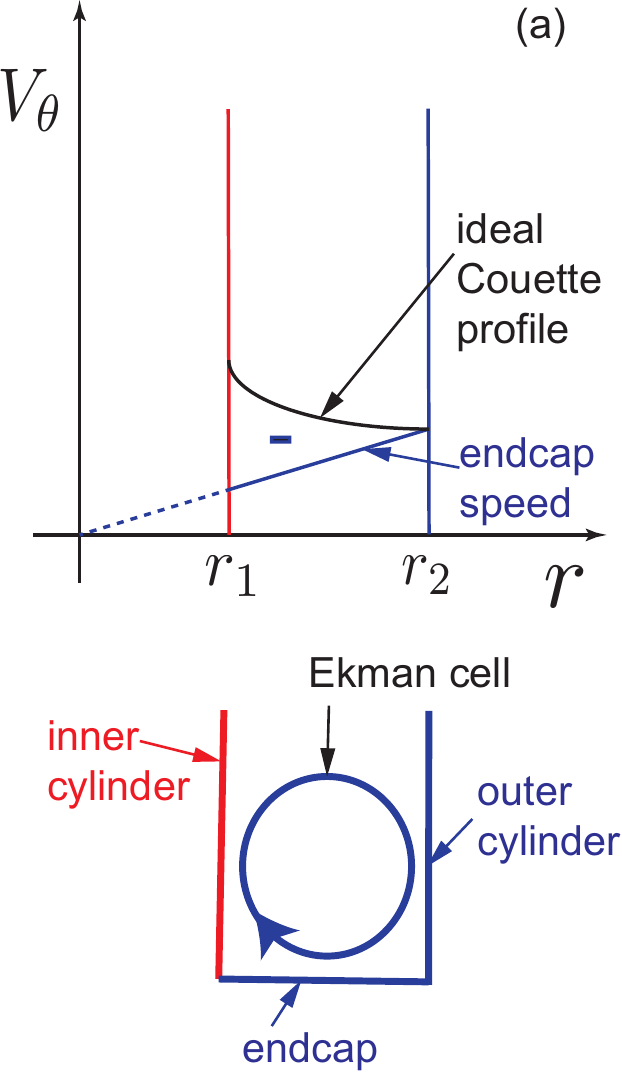}\hspace{0.5in}
\includegraphics[width=1.4in]{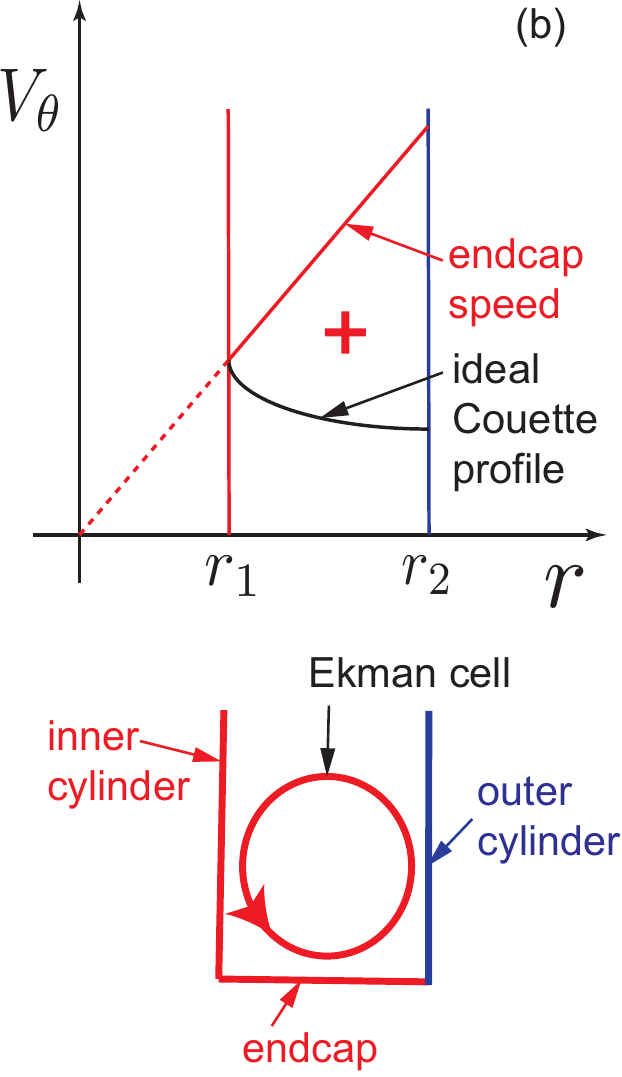}\hspace{0.5in}
\includegraphics[width=1.4in]{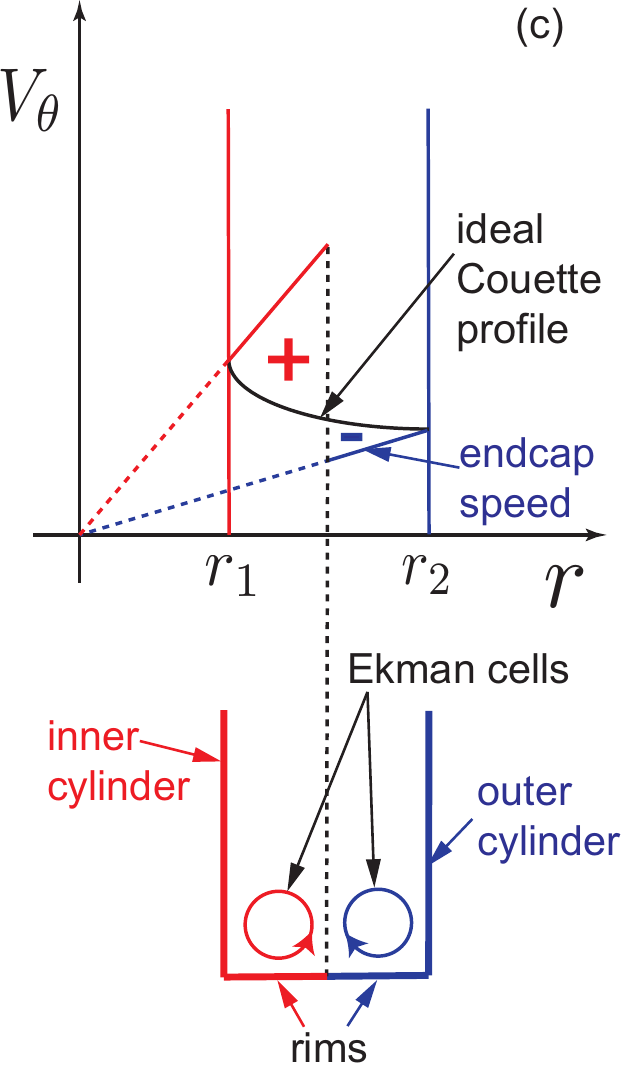}}\\
{\vspace{3mm}\\}
{\centering
\includegraphics[width=1.4in]{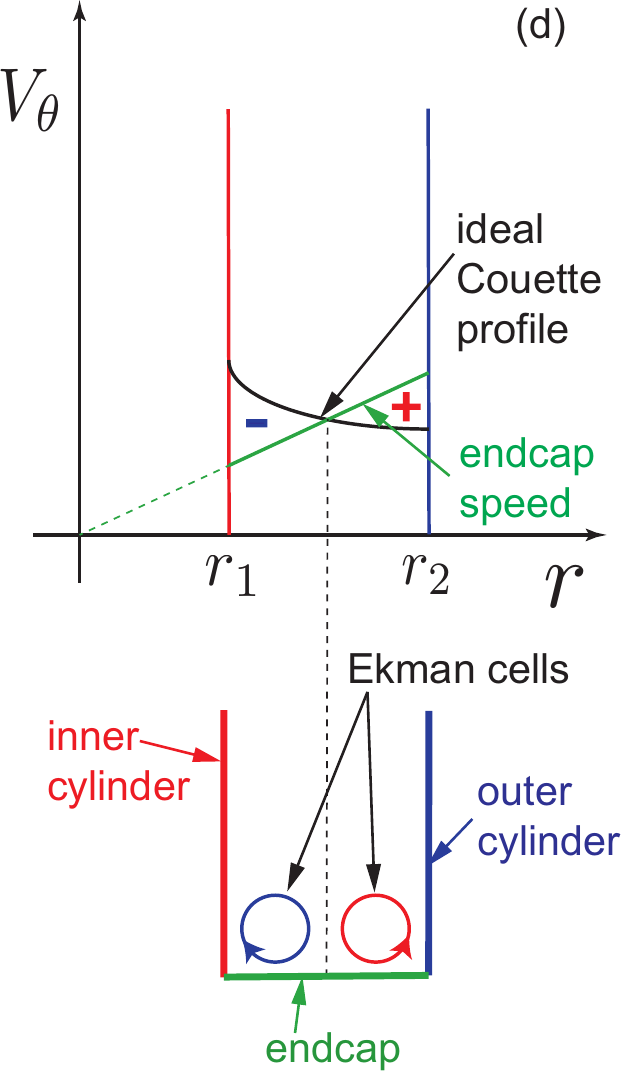}\hspace{0.5in}
\includegraphics[width=1.4in]{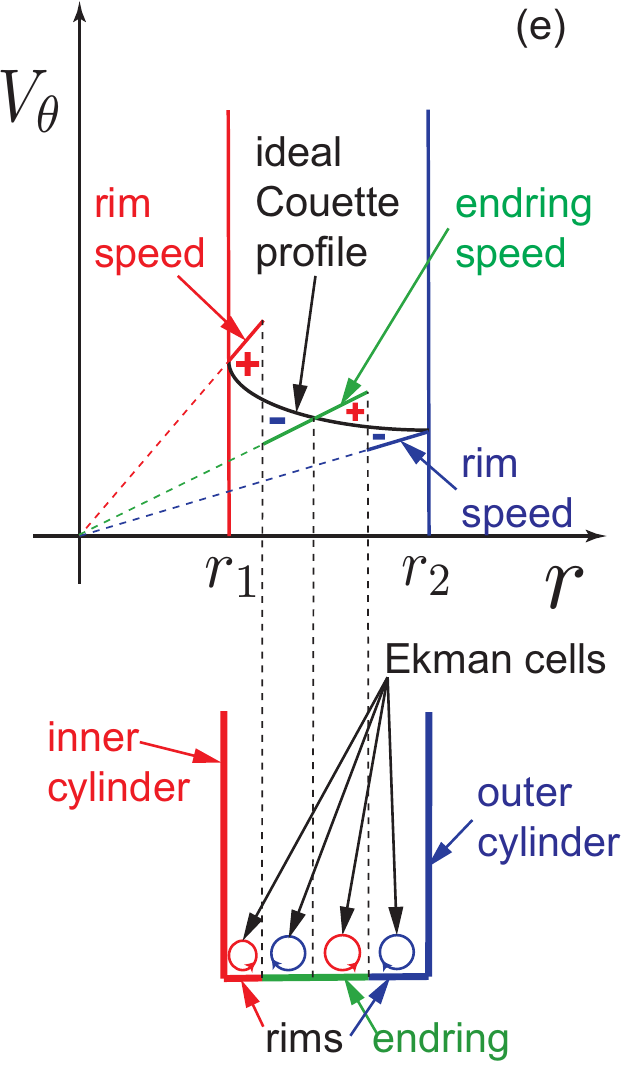}\hspace{0.5in}
\includegraphics[width=1.4in]{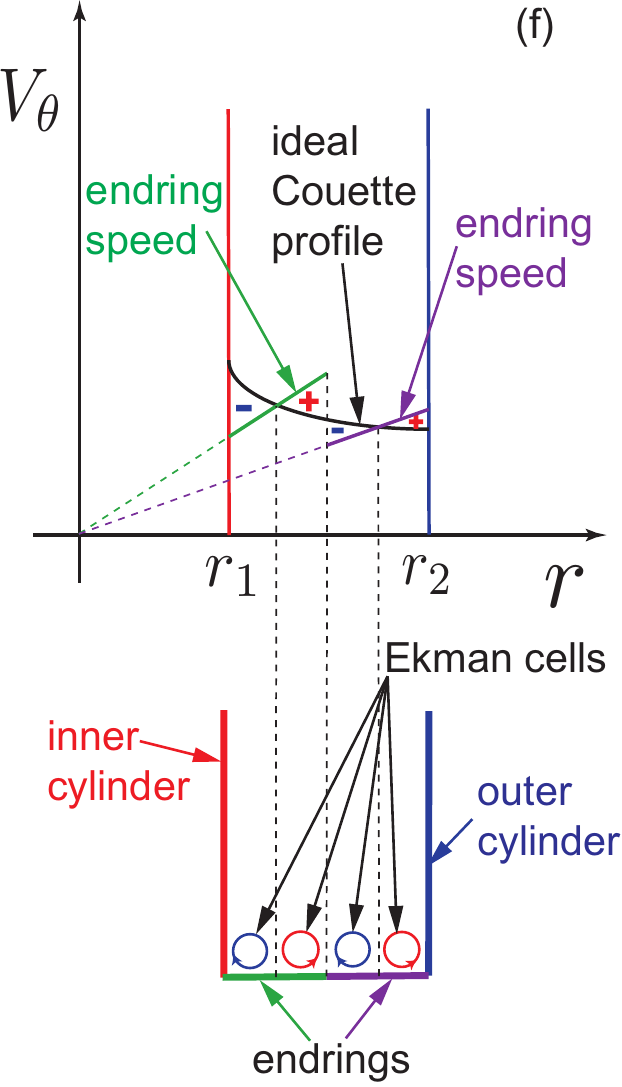}}
\caption{Illustrations of Ekman circulation induced by various endcap configurations. In each panel, the top figure shows color-coded velocity profiles of the endcaps, endrings, or rims. ``+" (``-") indicates speeds faster (slower) than the ideal-Couette profile.  The bottom figure illustrates Ekman cells in red (blue) induced by faster (slower) boundary rotation. (a) endcap (blue) attached to the outer cylinder (OC) \cite{wendt33,richard01,beckley02,kageyama04,paoletti11,paoletti12,nordsiek15}; (b) endcap (red) attached to the inner cylinder (IC)~\cite{wendt33}; (c) one rim (red) attached to IC and the other rim (blue) to OC (50/50 configuration~\cite{wendt33,richard01}); (d) one ring (green) spanning the radial gap and rotating independently of IC \& OC \cite{edlund15}; (e) as in (d) but with an inner rim (red) attached to IC and the other rim (blue) attached to OC~\cite{edlund14,edlund15}; (f) segmented endcap design with two independently rotating endrings (green and purple) spanning the radial gap without rims~\cite{ji06,schartman09,schartman12}. }
\label{fig_ekman}
\end{figure}

A particular challenge for quasi-keplerian TC studies is to minimize the effects of the axial boundaries, which would be stress-free in an astrophysical disc.
In principle one would have a very large aspect ratio $\Gamma\equiv h/(r_2-r_1)\gg 1$, but this becomes impractical when one needs a very large $Re$ based on the gap width $r_2-r_1$.
The importance of endcaps was recognized from the beginning when Mallock~\cite{mallock1896} used mercury as the bottom endcap in an attempt to reduce its effects (the top was a free surface). 
Endcaps corotating with the inner (outer) cylinder induce secondary Ekman circulation that flows radially outward (inward) near the endcaps, with the return flow in the bulk, as shown in Fig.~\ref{fig_ekman}(a-b). 
This circulation causes the primary azimuthal flow to deviate from the ideal-Couette profile, and vertical shear at the endcaps can produce turbulence at high enough $Re$. 
To reduce these effects, Taylor used $\Gamma\sim 100$~\cite{taylor1923,taylor36}, but the end effects still existed, as evidenced by the deviation from the ideal Couette profiles~\cite{taylor36b}. 

Wendt used a 50/50 split-endcap configuration~\cite{wendt33}, \textit{i.e.} the inner (outer) half of the endcap was attached to inner (outer) cylinder [see Fig.~\ref{fig_ekman}(c)], but this appeared to be insufficient at $\Gamma = 5-25$. Richard's experiment also employed split endcaps~\cite{richard01} at $\Gamma\approx 25$, but apparently the end effects were still significant (see \textit{e.g.} his Fig.4.2).
Therefore, it is unclear whether the observed turbulence was due to Ekman circulation or to the sought-after nonlinear radial-shear instability.

Endcaps effects were demonstrated by two subsequent experiments where endcaps were attached to the outer cylinder [Fig.~\ref{fig_ekman}(a)]. In \cite{beckley02}, the torque between the cylinders in a short TC ($\Gamma=2$) flow at $Re \sim 4\times 10^6$ was dominated by laminar Ekman effects on the endcaps rather than by turbulent transport in the bulk flow. Measurements of flow profiles and spin-down times in an even shorter TC flow ($\Gamma \approx 0.9$) were explained by scaled numerical modeling including the Ekman circulation~\cite{kageyama04}. 

\begin{figure}[ht]
{\centering\includegraphics[width=2in]{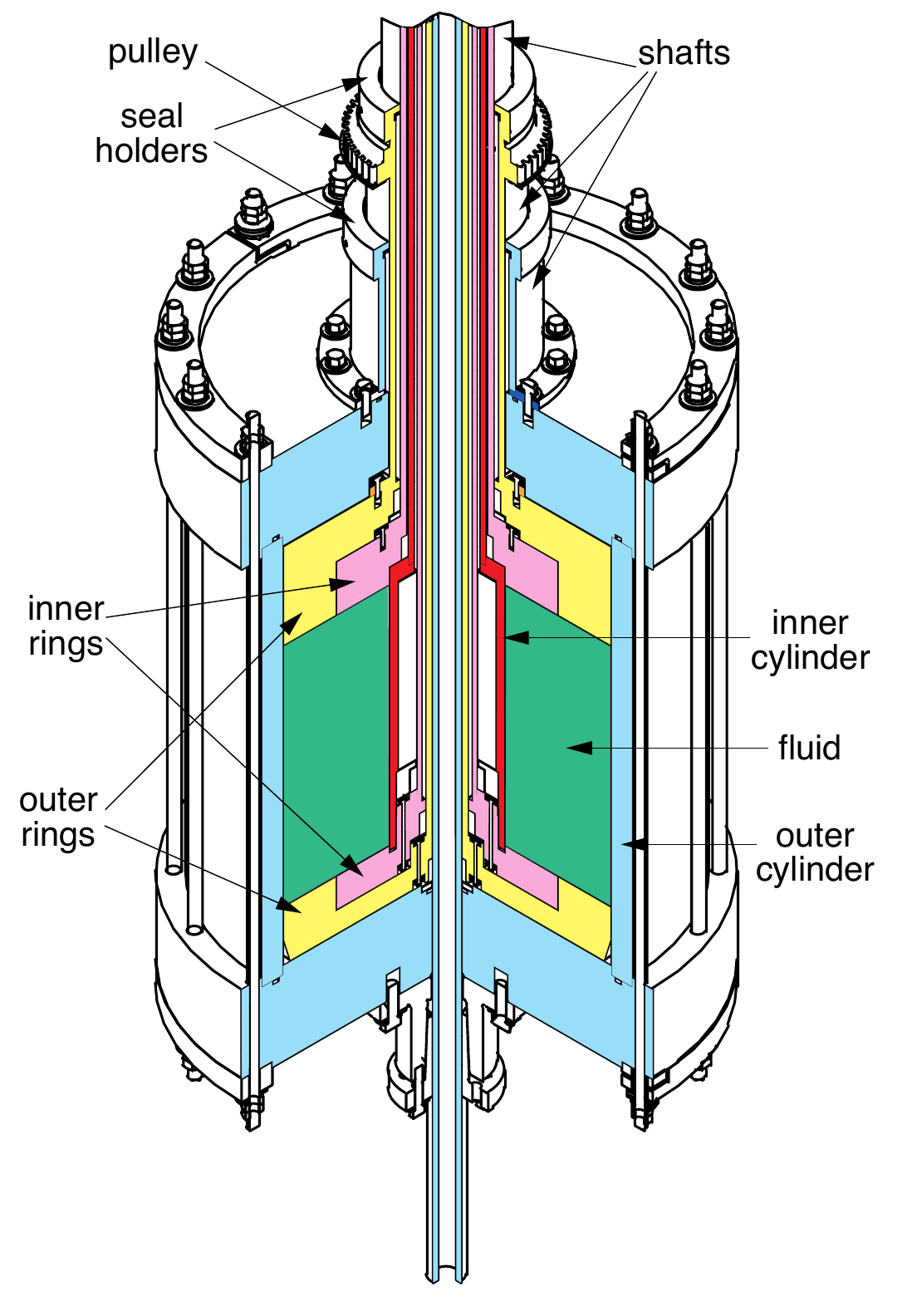}\includegraphics[width=3.3in]{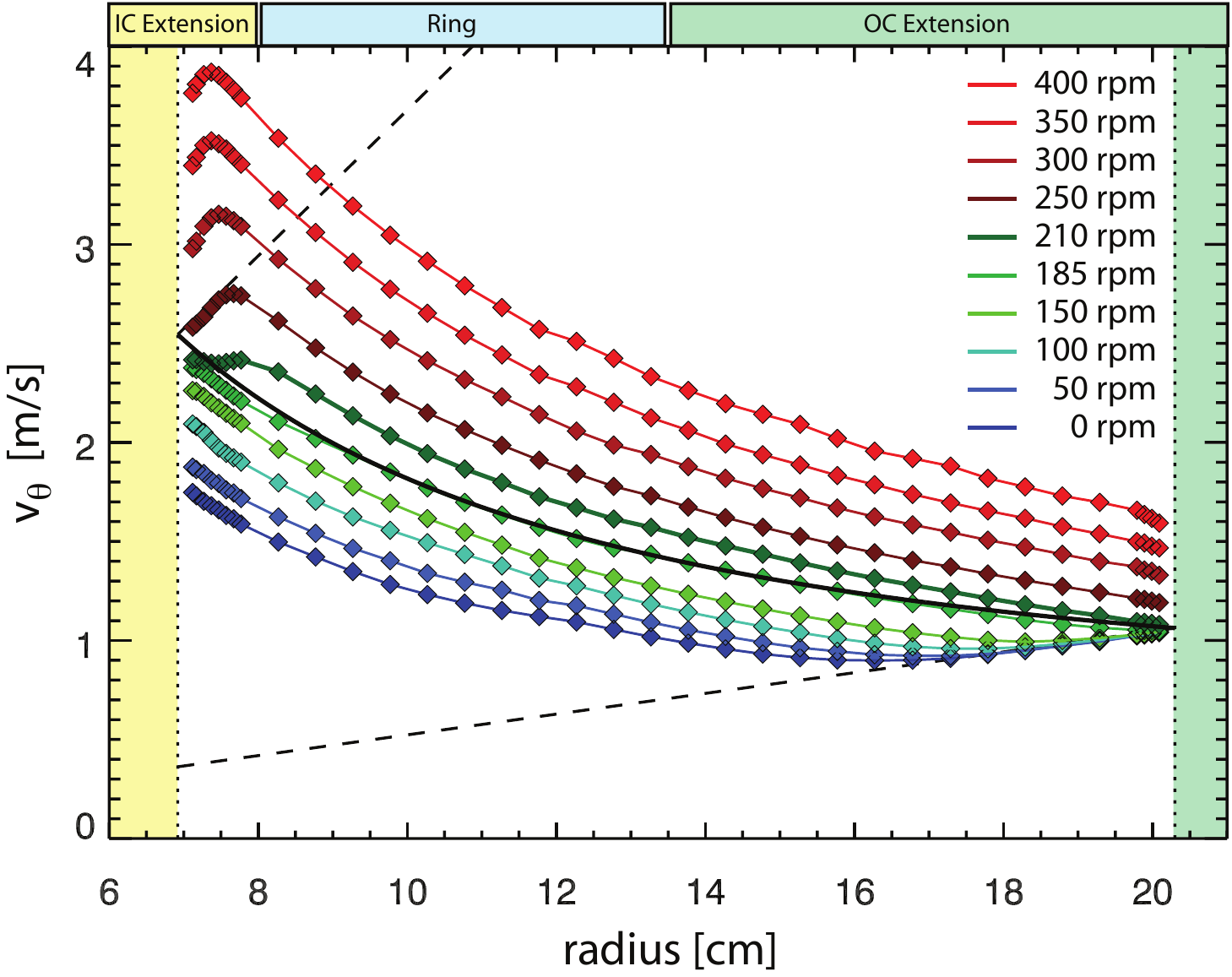}}
\caption{(a) Setup of Princeton MRI experiment during its hydrodynamic phase with two independently rotating endrings and aspect ratio $\Gamma\equiv h/(r_2-r_1)=2.1$. Adapted from Ref.\cite{ji06} (b) Measured velocity profiles in HTX ($\Gamma=2.96)$ at various ring speeds; 185 RPM yields a close match to the ideal-Couette profile (solid black line). Adapted from Ref.\cite{edlund14}. }
\label{fig_hydro}
\end{figure}

Based on this improved understanding, segmented endcap designs using multiple rings with independently controllable rotation speeds were explored via simulations~\cite{kageyama04,hollerbach04}.
The idea is to break up large Ekman cells that span the full radial gap and pervade the bulk flow into a number of smaller Ekman cells localized to the endcaps.
Evidently, the larger the number of small cells, the more closely the Ekman effect is confined to the endcaps. 
A specific design based on two endrings occupying the full radial gap [Fig.~\ref{fig_ekman}(f)] was implemented in the Princeton MRI experiment~\cite{ji01,goodman02}, as shown in Fig.~\ref{fig_hydro}(a).
Its effectiveness at eliminating Ekman effects in the bulk flow has been demonstrated by the fact that the ideal Couette profiles are restored by properly choosing the speeds of two endrings~\cite{burin06,schartman09,obabko08}.

A simpler design has been implemented in an experiment at Princeton called HTX (Hydrodynamic Turbulence Experiment)~\cite{edlund14}.
HTX has a single independently driven endring that is flanked by rims attached to each of the cylinders [Fig.~\ref{fig_ekman}(e)]. 
This is comparably effective to the MRI design at reducing Ekman circulation in the bulk flow, as shown by the azimuthal flow profiles in Fig.~\ref{fig_hydro}(b).
The circulation is divided, and thereby weakened, into the same number of Ekman cells as in the two-ring configuration [Fig.~\ref{fig_ekman}(f)].
Various endcap configurations are summarized in Fig.~\ref{fig_ekman}. The key is to increase the number of Ekman cells, while reducing their size, with minimum engineering complexity. 

After minimization of Ekman effects, quasi-keplerian flows are found to be essentially quiescent in the bulk as in solid-body flow, even at shear-based $Re\sim2\times 10^6$~\cite{ji06,schartman12}.
Furthermore, the flow reverts promptly to quiescence after perturbations (jets) as large as 20\% of the background flow are transiently imposed~\cite{edlund14}. 
This is consistent with numerical simulations for vertically periodic cylinders (hence without Ekman circulation):  turbulence decays as soon as the flows enter the quasi-keplerian regime from the centrifugally unstable one~\cite{ostilla14}, or after an optimal perturbation for transient growth is applied~\cite{shi17}. 

Reynolds stress and turbulent angular momentum flux have been directly measured by synchronized dual laser Doppler velocimetry, and found to be indistinguishable from zero for optimal endring speeds, at levels comparable to the viscous stress~\cite{ji06,schartman12}.
Conversely, with any non-optimal choices of endring speeds or when $Re$ is insufficiently large, large Reynolds stresses were detected.
These results are also consistent with direct numerical simulations showing that Ekman-driven turbulence penetrates into the bulk flow at low $Re$ even with optimal choices of endring speeds~\cite{avila12} while Ekman effects are confined to axial ends at large $Re$~\cite{lopez17}.

The profound effects of Ekman circulation particularly on quiescent quasi-keplerian flows offer a plausible explanation of the results reported by Paoletti et al. from an experiment at Maryland \cite{paoletti11,paoletti12}.
Significantly larger-than-laminar torques were measured on the middle section of the inner cylinder in an attempt to avoid Ekman effects near the endcaps. 
However, since the Maryland experiment had an aspect ratios of only $\Gamma=11.47$ and endcaps attached to the outer cylinder, significant Ekman effects are expected to penetrate well into the bulk flow~\cite{hollerbach04,avila12}. 
This was confirmed by velocity profiles measured in the Twente experiment, which had a similar geometry ($\Gamma=11.68$).
These show significant deviation from ideal-Couette profiles over a wide range of parameters in the quasi-keplerian regime~\cite{nordsiek15}.
High-shear flow near the inner cylinder was found to transport angular momentum both radially outward and axially towards endcaps~\cite{nordsiek15}. 

Although the nonlinear stability of quasi-keplerian TC flows seems to be settled empirically up to $Re\sim 10^6$, there remain several critical unanswered questions:
\begin{itemize}
    \item Astrophysical Reynolds numbers are huge: for example, in the so-called Minimum-Mass Solar Nebula \cite{Hayashi1981} $Re\equiv |S| h^2/\nu\approx 10^{13}(r/r_\oplus)^{-3/2}$, $h$ being the half thickness of the disc, $r_\oplus\approx 1.5\times10^{11}\unit{m}$ the Earth's orbital radius, and $S=rd\Omega/dr=(-3/2)\Omega$ the radial shear.
    Perhaps quasi-keplerian flow becomes turbulent at $Re\gg 10^6$ but $<10^{13}$?
    Experience with subcritically unstable shear flows (e.g. pipe flow~\cite{mullin2011experimental}) suggests, however, that turbulent momentum transport declines with increasing $Re$ when scaled by quantities independent of viscosity.
    Hence if hydrodynamic disc turbulence exists, it may be feeble.
    \item Quasi-keplerian \emph{compressible} plane-Couette flow is linearly unstable when the shear is supersonic \cite{Narayan+1987}.
    However, instability requires at least one highly reflecting boundary, and peak growth rates are $\sim 10^{-3}|S|$ at keplerian ratios of shear to rotation ($-3/2$).
    So the importance of these instabilities for accretion is doubtful, except perhaps in the boundary layer where the disk meets its star \cite{Belyaev+Rafikov2012}.
    \item There can be hydrodynamic disc instabilities driven by vertical (i.e., parallel to the rotation axis) thermal gradients~\cite{Lesur+2022}.
    However, short cooling times are required for rapid growth, and the resulting turbulence tends to be weak and small-scale. More importantly, whether such turbulence can indeed effectively transfer angular momentum \textit{radially}, as demanded in astrophysics, still remains to be seen.
    \item Can we achieve larger $Re$ experimentally using liquid He or compressed gas?
    \item Can we achieve larger $Re$ numerically using subgrid models etc?
    \item Most ambitiously, can the nonlinear hydrodynamic stability of quasi-keplerian flows be proven mathematically?
    The second and third bullets above suggest that the conditions of any such theorem will need to be carefully formulated.
\end{itemize}
        
\section{Magnetohydrodynamic Experiments Using Liquid Metals}

\begin{figure}[ht]
\vspace{-1cm}
\centering\includegraphics[clip, trim=0cm 1cm 9cm 2cm, width=2in]{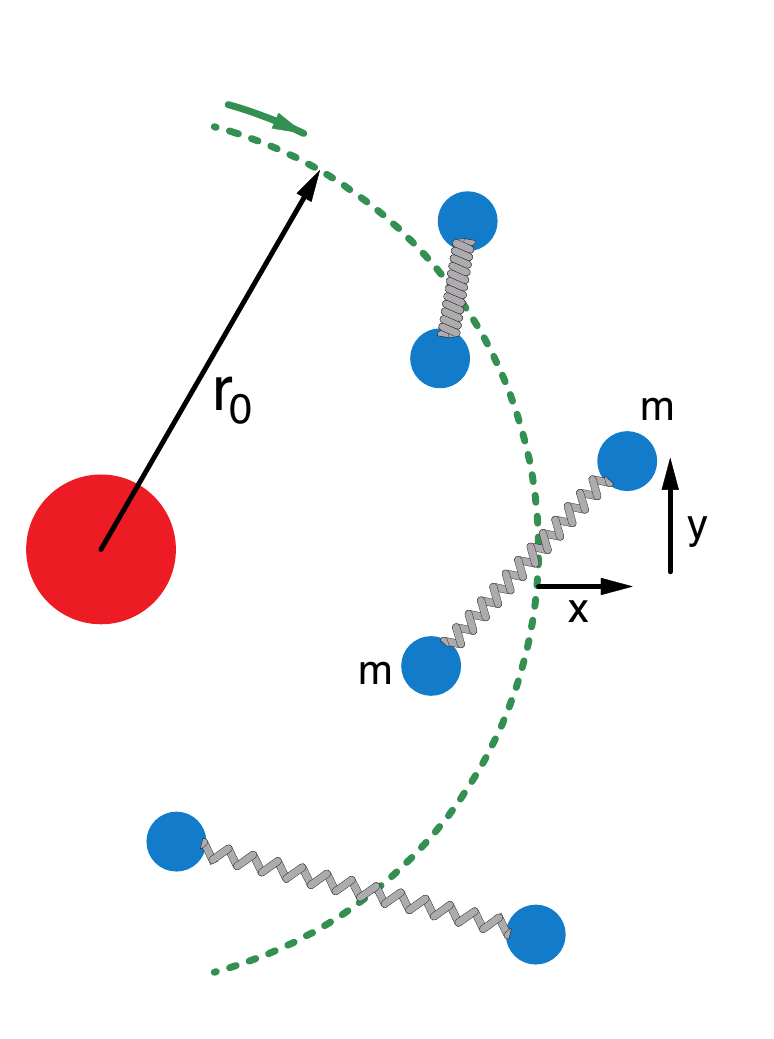}
\caption{A spring-mass analogue of standard magnetorotational instability (SMRI) originally suggested by A.~Toomre (unpublished) and elaborated in terms of a local dispersion relation~\cite{balbus92a}.
Satellites or masses (solid blue circles) of a central gravitating mass (red) follow near-circular orbits with angular frequencies $\Omega(r)$ while tethered by a weak harmonic spring (gray) having squared natural frequency $\omega_0^2<-rd\Omega^2/dr$.
The satellites tend to separate!  Conversely, a weak repulsion keeps them together.
Adapted from Ref.\cite{hung19}. }
\label{fig_spring}
\end{figure}

\begin{table}[ht]
\caption{Magnetohydrodynamic TC-flow experiments using liquid metal on Standard Magnetorotational Instability (SMRI), Helical MRI (HMRI), and Azimuthal MRI (AMRI) and relevant direct numerical modeling.}
\label{MHD_experiments}
\begin{tabular}{lp{2.9cm}p{7.1cm}p{1.2cm}}
\hline
Year & Author(s) & Main results & Exp or modeling \\
\hline
2001 & Ji et al.~\cite{ji01}; Goodman \& Ji~\cite{goodman02}; R\"{u}diger \& Zhang~\cite{ruediger01} & Proposed realization of SMRI in liquid-metal, quasi-keplerian TC flow with axial magnetic field & Modeling \\
2004 & Sisan et al.~\cite{sisan04} & Claimed detection of SMRI in spherical Couette flow but it turned out to be Shercliff-layer instability & Exp \\
2005 & Hollerbach~\& R\"{u}diger \cite{hollerbach05} & Predicted HMRI by adding an azimuthal magnetic field with a much reduced critical $Rm$ & Modeling\\
2006 & Liu et al.~\cite{liu06a,liu08a} & Predicted nonlinear saturation of SMRI & Modeling\\
2006 & Liu et al.~\cite{liu06b,liu07} & Identified HMRI as a weakly destabilized, traveling inertial oscillation, stable in keplerian flows & Modeling\\
2006 & Stefani~\cite{stefani06,stefani09} & HMRI detected, also with reduced Ekman effects & Exp\\
2007 & Priede et al.~\cite{priede07} & Inductionless nature of HMRI & Modeling \\
2007 & R\"{u}diger et al.~\cite{ruediger07b}; Hollerbach et al.~\cite{hollerbach10} & Predicted inductionless AMRI with pure azimuthal magnetic field with much lower critical $Rm$ & Modeling\\
2009 & Boldyrev et al.~\cite{boldyrev09}; Bai et al.~\cite{bai15,bai21} & Viscoelastic instability in polymer fluids as an SMRI analogue & Exp\\
2010 & Nornberg et al.~\cite{nornberg10} & Shercliff layer instability measured on outer cylinder, interpreted as magnetocoriolis waves & Exp\\
2011 & Gissinger et al.~\cite{gissinger11} & Identified the instability observed in spherical Couette flows as Shercliff layer instability & Modeling\\
2012 & Roach et al.~\cite{roach12,roach13}; Spence et al.~\cite{spence12} & Free Stewartson-Shercliff layer instability detected via internal flow measurements and modeling & exp and modeling\\
2012 & Gissinger et al.~\cite{gissinger12} & Imperfect bifurcation to SMRI due to Ekman effects & Modeling\\
2014 & Seilmayer et al.~\cite{seilmayer14} & AMRI detected & Exp\\
2016 & Wei et al.~\cite{wei16}; Winarto et al.~\cite{winarto20} & Effects of conducting endcaps on SMRI: larger amplitude with lower critical $Rm$ & Modeling\\
2018 & Caspary et al.~\cite{caspary18}; Choi et al.~\cite{choi19} & Effects of conducting endcaps on free Shercliff layer instability & Exp and Modeling \\
2019 & Hung et al.~\cite{hung19} & Spring-mass analogue of SMRI in quasi-keplerian flows & Exp\\
2021 & Vernet et al.~\cite{vernet21,vernet22} & Keplerian turbulence driven by radial current  & Exp\\
2022 & Wang et al.~\cite{wang22a,wang22b} & First observation of axisymmetric SMRI and a new non-axisymmetric SMRI-like mode & Exp and modeling\\
\hline
\end{tabular}
\vspace*{-4pt}
\end{table}

We turn now to TC experiments in which the working fluid is an electrical conductor, usually a liquid metal, and on which a background magnetic field is imposed with axial ($B_z$) or azimuthal ($B_\theta$) components, or both.
Room-temperature liquid metals are rather poor conductors compared to silver, copper, or aluminum, so the magnetic diffusivity $\eta=1/\mu\sigma$, which has the same units as kinematic viscosity, is large and important for such experiments.
(Here $\mu\approx\mu_0$ is the magnetic permeability and $\sigma$ the conductivity of the fluid.)
As in hydrodynamics, the flow parameters are best expressed in dimensionless combinations.
With $\eta$ and $B\equiv\sqrt{B_z^2+B_\theta^2}$, one can form two new independent dimensionless parameters in addition to $Re$.
The magnetic Prandtl number $Pm\equiv\nu/\eta$ is a pure material property, typically $10^{-5}$ to $10^{-7}$ in liquid metals.
The Lehnert number $Le\equiv V_A/r_1\Omega_1$ compares the Alfv\'en speed $V_A=B/\sqrt{\mu\rho}$ (the speed of compressionless waves restored by magnetic tension) to the rotation speed, $\rho$ being the fluid density.
Alternative dimensionless combinations are often used for various ratios of forces or timescales.
These include magnetic Reynolds number $Rm = r_1^2\Omega_1/\eta = Pm Re$, which compares the timescales of magnetic diffusion and rotation; Lundquist number $Lu = r_1 V_A/\eta = Rm Le$; Hartmann number $Ha = r_1V_A/\sqrt{\nu\eta} = Le Re Pm^{1/2}$; and Elsasser number $\Lambda = V_A^2/\eta\Omega_1 = Le^2 Rm$.
Since $Pm\lesssim 10^{-5}$, the regime $Rm>1$, which is necessary for the astrophysically important standard magnetorotational instability (SMRI: see below) can be reached only at $Re\gtrsim 10^5$.
However, SMRI has interesting cousins (HMRI, AMRI; also see below) that can be unstable in the limit $Rm\to0$ provided $Ha$ and $\Lambda$ are large enough.

The history of SMRI traces back to late 1950s~\cite{velikhov59,chandra60}, but it was not proposed until early 1990s as the source of accretion-disc turbulence~\cite{balbus91,balbus98}.
The mechanism of SMRI is illustrated by a spring-mass analogue shown in Fig.~\ref{fig_spring}. 
The tethered masses exchange angular momentum via tension in the spring.
Because of the opposing signs of $d\Omega/dr$ and $d(r^2\Omega)/dr$, the mass at lower altitude moves ahead in azimuth, progressively losing angular momentum and altitude while increasing its angular velocity; the reverse holds for the other mass.
This can be also understood as the increased (decreased) radial force on the inner (outer) mass can overcome the spring force to pull (push) the inner (outer) mass further away from their original positions, leading to a runaway instability.
Therefore this happens only if the spring is sufficiently weak.
If the spring is too strong, the perturbations are stabilized (oscillatory). 

In SMRI, the role of the spring is played by magnetic tension, exchanging angular momentum between fluid elements on the same field line.
In the simple case that the background field is uniform and axial (\emph{i.e.}, parallel to the rotation axis), and in ideal MHD ($\eta=\nu=0$), the dispersion relation for the complex angular frequency $\omega$ reads
\begin{equation}\label{eq_SMRI}
\omega^4 - \left(2\omega_A^2 + \kappa^2\frac{k_z^2}{k^2}\right) \omega^2
+\omega_A^2 \left( \omega_A^2 + r \frac{d\Omega^2}{dr}\frac{k_z^2}{k^2}\right) =0,
\end{equation}
in which $\omega_A^2\equiv \left(\boldsymbol{k} \cdot \boldsymbol{V}_A\right)^2=k_z^2 B_z^2/\mu\rho$ is in effect the square of the natural frequency of the ``spring," $k=\sqrt{k_r^2+k_z^2}$ is the total wavenumber for axisymmetric modes, and  $\kappa^2 \equiv r^{-3}d(r^4\Omega^2)/dr$ is the square of the epicyclic frequency.
Note $\kappa^2<0$ and $\omega_A=0$ yields Rayleigh's centrifugal instability.
In the quasi-keplerian regime, $\kappa^2>0$ and $d\Omega^2/dr<0$, instability ($\mathrm{Im}\,\omega>0$) exists when $0< \omega_A^2<-rd\Omega^2/dr$ so that the last time in Eq.\eqref{eq_SMRI} is negative.
In non-ideal MHD, with shifted frequencies $\omega_\nu\equiv\omega+i\nu k^2$ \& $\omega_\eta\equiv\omega+i\eta k^2$, one has (e.g. \cite{pessah08})
\begin{equation}\label{eq_nonideal}
    (\omega_\nu\omega_\eta-\omega_A^2)^2+\left[-\kappa^2\omega_\eta^2 + \omega_A^2 r\frac{d\Omega^2}{dr}\right]\frac{k_z^2}{k^2}=0\,.
\end{equation}
For $\nu>0$ and $\eta=0$ ($Pm=\infty$), a marginal mode ($\omega=i0^+$) always exists.
But in the opposite limit ($0\le\nu\ll\eta$), which is relevant for liquid metals, SMRI requires a threshold $Rm$ on the order of unity based on the marginal stability condition $\eta k^2\kappa>(-rd\Omega^2/dr)/2$ for optimal $\omega_A^2$ and $\kappa^2>0$. 

Experimentally in TC flows, the spring-mass analogue has been successfully demonstrated~\cite{hung19} using an embedded ball in quasi-keplerian regime. The ball motion in terms of local Lagrangian displacement $x$ and $y$ (see Fig.~\ref{fig_spring} for definitions) follows predictions by the same dispersion relation Eq.\eqref{eq_SMRI} when $\omega_A^2$ is replaced by the spring constant and $k_r=0$ with other minor modifications. Similarly, another analogue of SMRI by viscoelastic instability in polymer fluids where long and elastic polymers act like magnetic field in the linear regime has been verified experimentally~\cite{boldyrev09,bai15,bai21}.

In contrast, the search for SMRI in liquid-metal TC flow has been long and arduous, fundamentally because $Pm\ll 1$.
This drives one to use wide radial gaps in order to minimize $\eta k_r^2$, leading in turn to small aspect ratios and the attendant end effects.
Table~\ref{MHD_experiments} lists TC flow experiments in liquid metals and relevant direct numerical modeling in quasi-keplerian regime, including the aforementioned MRI analogues.

The first liquid-metal TC experiment~\cite{donnelly60} was performed using mercury in the Rayleigh-unstable regime (see Fig.~\ref{fig_diagram}). This was motivated by---and actually confirmed---the theoretical prediction that, contrary to MRI, adding magnetic field stabilizes Rayleigh instability, leading to increased critical $Re$ with magnetic field strength~\cite{chandra60}.  This can be understood from Eq.~\eqref{eq_nonideal}.

\begin{figure}[!ht]
{\centering\includegraphics[width=3in]{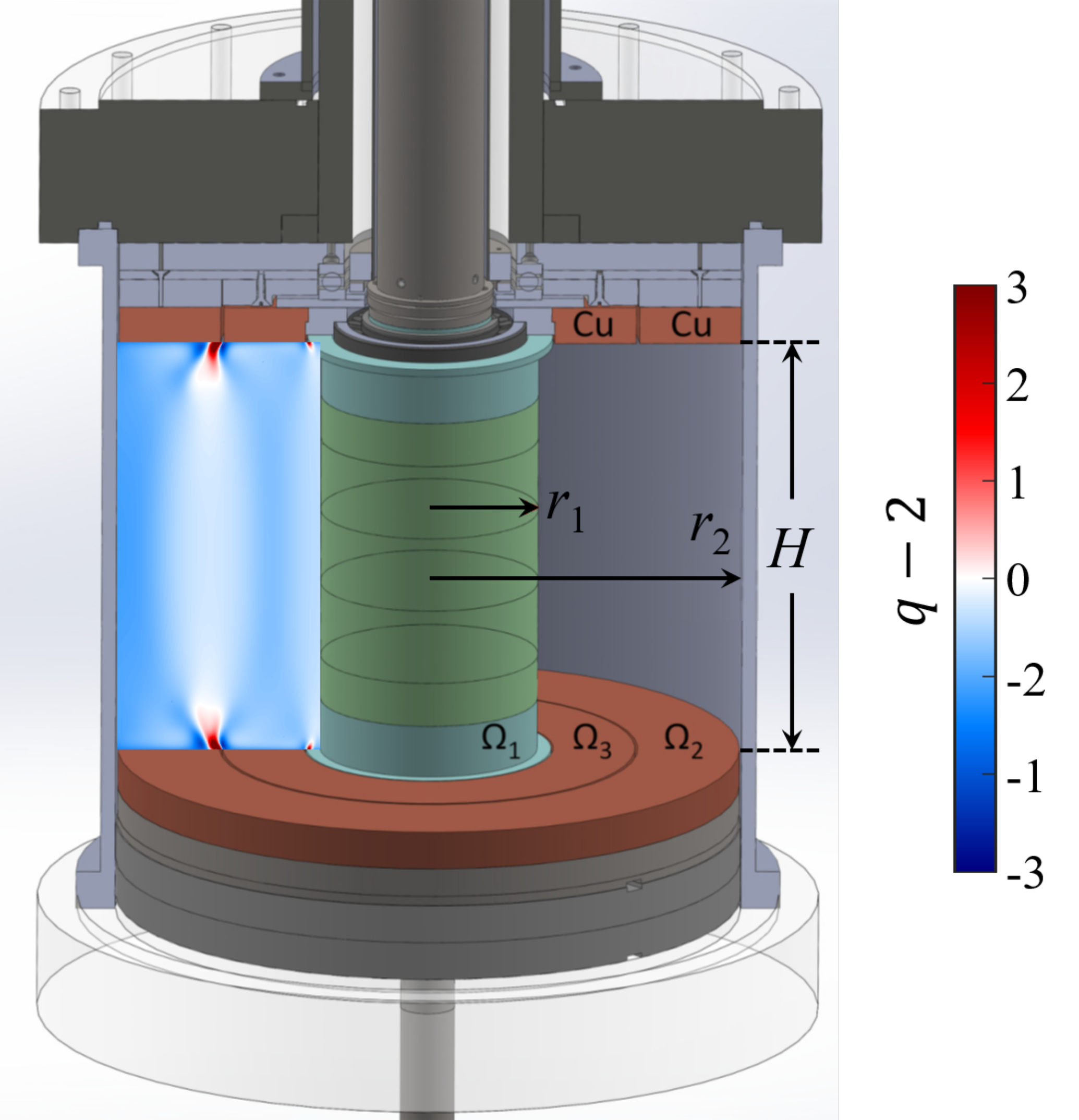}\hspace{2mm}\includegraphics[width=2.2in]{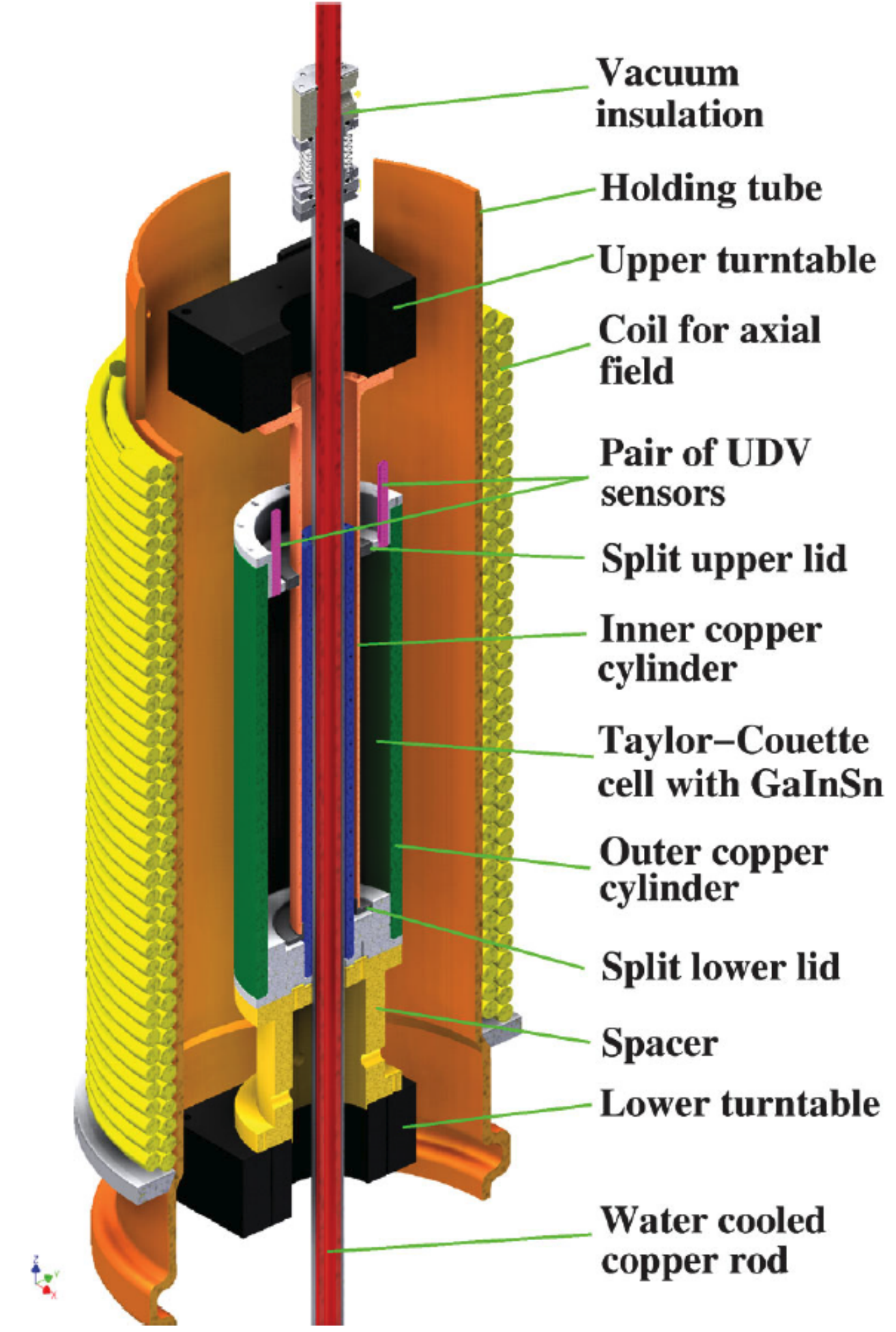}}
\caption{(Left) The Princeton MRI apparatus with conducting rotating endcaps.
Overlaid on the left is the azimuthally averaged shear profile, $q-2\equiv -d\log(r^2\Omega)/d\log r$, in a 3D hydrodynamic simulation.
Adapted from \cite{wang22a}. 
(Right) The PROMISE apparatus, supporting both vertical and azimuthal magnetic field. Adapted from \cite{seilmayer14}. }
\label{fig_MHD}
\end{figure}

Liquid-metal TC experiments attracted limited interest before Balbus \& Hawley's theoretical rediscovery of MRI in the early 1990s, and even then only after a proposal to study SMRI in a quasi-keplerian liquid-gallium TC apparatus~\cite{ji01,goodman02}. 
The dependence on $Pm$ for such experiments was analyzed \cite{ruediger01,willis02}, and the feasibility of a liquid-sodium experiment was also shown~\cite{noguchi02}.
Nonlinear saturation~\cite{knobloch05,liu06a,Umurhan+2007,liu08a}, and even dynamo action~\cite{willis02b} were explored through asymptotic analysis (in narrow gaps, excepting \cite{clark+oishi17}) and numerical simulation.
Experimentally, however, SMRI was not positively detected until very recently, another 20 years later~\cite{wang22a,wang22b}, and only after the proper vertical boundary conditions on endcaps were first studied numerically~\cite{wei16,winarto20}, then implemented and confirmed experimentally~\cite{caspary18,choi19}.

Endcaps complicate a magnetized liquid-metal TC flow, but can also be exploited to advantage. The complication comes from the enhanced Ekman/Hartmann effects. A discontinuity in the angular velocity at the axial boundary can generate a free Stewartson-Shercliff layer or SSL~\cite{stewartson57,shercliff53,lehnert55} extending vertically into the bulk flow and reinforced by the applied $B_z$.
The layers can be unstable to nonaxisymmetric Kelvin-Helmholtz instabilities~\cite{roach12,spence12,roach13,nornberg10,caspary18,choi19}. The magnetized version of SSL instability is inductionless (see below) and occurs when the Elsasser number $\Lambda >1$.
The unstable free SSLs formed tangential to inner sphere of spherical Couette flows~\cite{hollerbach01} also explain~\cite{gissinger11} a mistaken claim to detect MRI~\cite{sisan04}.

The advantage of the endcaps in liquid metal TC experiments comes from the reinforcement of SMRI-unstable rotation via \textit{electrically conducting} endcaps, allowing SMRI to saturate at more detectable amplitudes. 
In fact, the first attempts to detect SMRI using insulating endcaps failed even though the required rotation speed and magnetic field strength were achieved~\cite{roach13}. 
This was due to the weak viscous coupling between liquid metal and \textit{insulating} boundaries at $Re\gtrsim 10^6$, so that saturation occurred undetectably small modifications of the flow and field \cite{knobloch05}. To increase the coupling, conducting endcaps using copper were proposed~\cite{wei16}, implemented (see Fig.~\ref{fig_MHD}, left panel) and shown to be effective~\cite{caspary18}. 

\begin{figure}[!ht]
{\centering\includegraphics[width=2.7in]{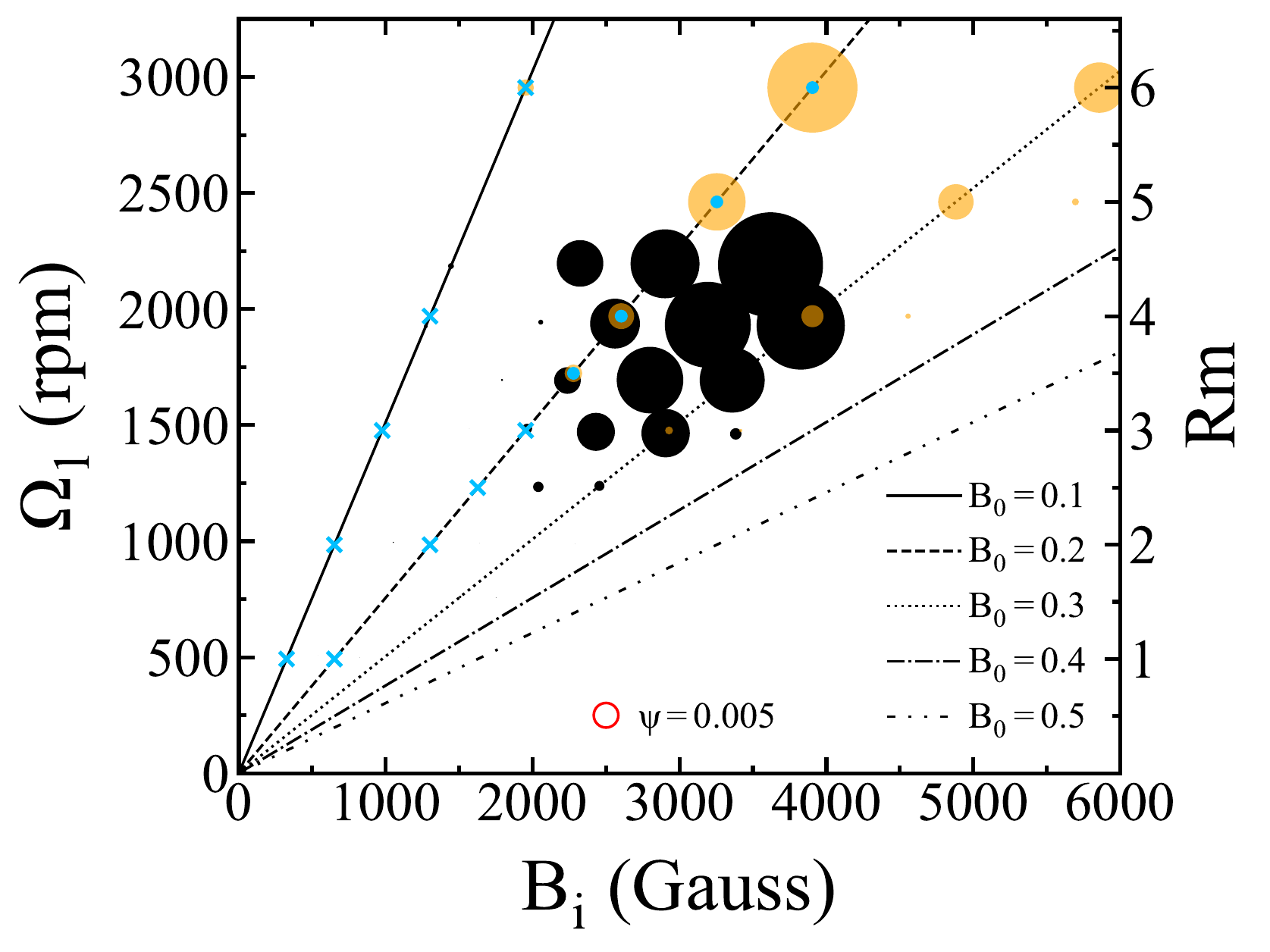}\hspace{2mm}\includegraphics[width=2.7in]{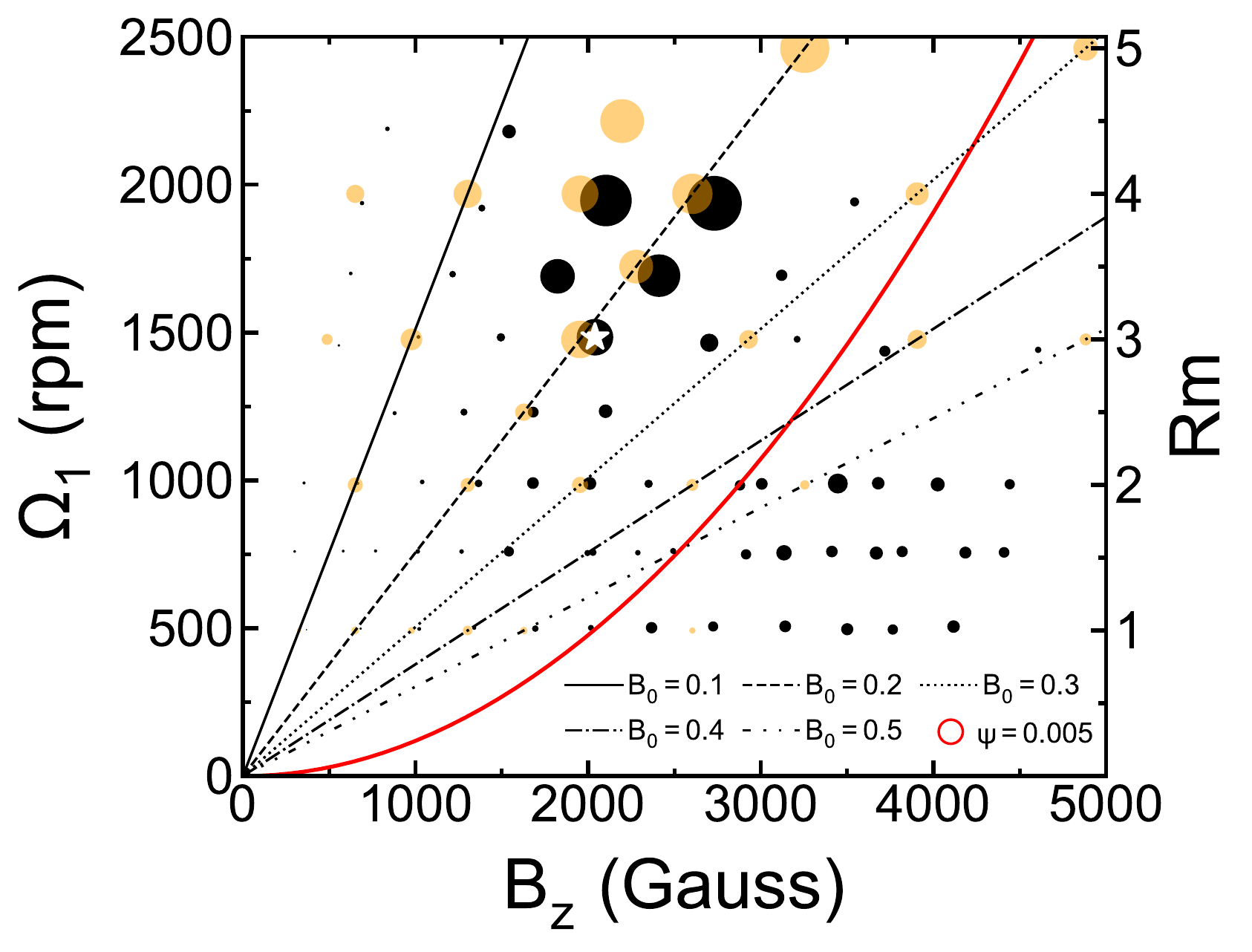}}
\caption{(Left) ``Bubble plot" of axisymmetric SMRI amplitude from Princeton MRI experiments (black) and 3D simulations (orange).
Bubble diameter is proportional to $B_r$ on the inner cylinder at $z=H/4$, halfway between midplane \& upper endcap. Blue crosses and dots show predictions from global linear analysis that the $m=0$ mode is stable (crosses) or unstable (dots) in a $q<2$ average bulk profile $\overline{V}_\theta(r)$ from a 3D MHD simulation run to saturation. $B_0=Le=B_z/r_1\Omega_1\sqrt{\mu\rho}$ is the Lehnert number. Adapted from \cite{wang22a}.
(Right) ``Bubble plot" of $m=1$ amplitude from experiments (black) and 3D simulations (orange). Bubble diameter is proportional to $m=1$ radial magnetic field on the inner cylinder at the midplane. 
Red curve represents the boundary for Stewartson-Shercliff layer instability. Adapted from \cite{wang22b}. Note that the bubble size has been adjusted so that they are on the same scale between two panels. The red circles show 0.5\% of the imposed field.}
\label{fig_SMRI}
\end{figure}

However, the magnetically enhanced coupling to the flow via conducting  endcaps accentuates the SSL and makes approximating ideal-Couette profiles harder even with segmented endcaps (\S2). 
Radial magnetic field, $B_r$, the primary diagnostics of SMRI activity, can also be induced by residual Ekman circulation even before the onset of SMRI. 
Since $B_r$ due to residual Ekman circulation is proportional to the imposed $B_z$, the concept of ``imperfect bifurcation" has been introduced~\cite{gissinger12,knobloch96}.
Increase of the proportionality $B_r/B_z$ with increasing $Rm$ or $Le$ can be used to identify the SMRI onset~\cite{wei16}. 
Because of the imperfect bifurcation, the critical $Rm$ is lowered substantially~\cite{winarto20} from the linear prediction for ideal Couette profiles, as an additional benefit of conducting endcaps. An exponential growth phase does not occur because the magnetized flow does not start from a steady state; more precisely, it starts from a hydrodynamic, but not MHD, steady state. Instead, the flow evolves into a new state that is distinctly affected by MRI at large enough $Rm$. The saturated
$B_r/B_z$ amplitude due to SMRI is shown in Fig.~\ref{fig_SMRI} (left) in the $\Omega_1-B_z$ diagram, after removing the contribution from residual Ekman circulation from the measured $B_r/B_z$. It compares well with data from 3D modeling following the same procedure. 

The identified instability exhibits several characteristics of SMRI: unstable only above a critical $Rm$, and stable when $B_z$ is too small due to finite dissipation and also when $B_z$ is too strong, \textit{c.f.} the dispersion relations \eqref{eq_SMRI}-\eqref{eq_nonideal}. 
The case for SMRI is further supported by global linear analysis of averaged flow profiles over the simulated bulk flow, which is in the quasi-keplerian regime, \textit{i.e.} $q<2$, see Fig.~\ref{fig_MHD}(left).
The $B_r/B_z$ signal becomes large where SMRI is predicted.
While these characteristics are consistent with expectations for $m=0$ SMRI, other explanations cannot be ruled out; further study is called for.

However, there is a surprise. The detected SMRI is axisymmetric (azimuthal mode number $m=0$), as is predicted to be unstable at the lowest $Rm$ threshold. 
But in addition, linearly growing nonaxisymmetric modes, especially $m=1$ modes, are also detected~\cite{wang22b}, and their saturated amplitudes are shown in Fig.~\ref{fig_SMRI}(right) in the $\Omega_1-B_z$ diagram. A subtle but important distinction is that these nonaxisymmetric modes are largely $z$-\textit{independent}, while axisymmetric SMRI modes are $z$-\textit{dependent} due to their two-cell structure. To place them in their proper prospective, the mode amplitudes are shown in Fig.~\ref{fig_SMRI} based on the same scale, indicating that the $m=1$ amplitude is at least a factor of 2 less than the $m=0$ SMRI amplitude measured at $z=H/4$. 

These nonaxisymmetric mode observations cannot be explained by residual hydrodynamic modes, nor by the SSL instabilities, which exist only to the right of the red curve~\cite{wang22b}. 
They may be related to nonaxisymmetric modes found in the narrow-gap limit by Ref.~\cite{oishi20}, but they occur at $Rm$ smaller by a factor 8-10 than what would be predicted for an ideal-Couette profile. 
The fact that the $m=1$ modes are observed in nearly the identical parameter space as the $m=0$ SMRI modes suggests that they are related. 
Further study is needed to understand the nonaxisymmetric modes.  Nevertheless, they have been reproduced numerically in 3D simulations, where they grow exponentially upon the (SMRI-modified) axisymmetric background state, and hence are true linear instabilities~\cite{wang22b}.

The experimental search for SMRI in liquid-metal TC flows has led to related but inductionless instabilities: Helical MRI (HMRI)~\cite{hollerbach05,stefani06,liu06b,stefani09} and Azimuthal MRI (AMRI)~\cite{ruediger07,hollerbach10,seilmayer14}.
The former involves an imposed field with both azimuthal ($B_\theta$) and axial ($B_z$) components; the latter, $B_\theta$ only. 
Due to the much lower $Re$, Ekman effects are important~\cite{stefani09}. HMRI is essentially a weakly destabilized $m=0$ inertial oscillations propagating axially~\cite{liu06b}, while $m=1$ AMRI modes propagate azimuthally. 
Both were first predicted theoretically~\cite{hollerbach05,hollerbach10} and later confirmed in the PROMISE experiment \cite{stefani06,seilmayer14}: see Fig.~\ref{fig_MHD}(right). 

Of interest also is a recent novel liquid-metal TC experiment called KEPLER that has a very small aspect ratio, $h\ll r_1,r_2$ \cite{vernet21}.
Confined vertically by insulating (Plexiglas) endplates, and radially by stationary conducting cylinders held at different voltages, the azimuthal flow is driven by a radial current and an applied axial magnetic field, a scheme proposed previously to study SMRI~\cite{velikhov06b}.
The balance between the applied $\boldsymbol{J\times B}$ force and turbulent drag against the endplates results in a keplerian profile, $V_\theta\propto r^{-1/2}$.
Thus the KEPLER experiment resembles a thin astrophysical disc in its flow profile as well as its geometry.
Although the flow transports angular momentum outward \cite{vernet22}, the source of its turbulence is probably not any version of MRI, but rather interaction with the endplates.

HMRI, AMRI, and the SSL instability are inductionless.
By this we mean the magnetic perturbation $\delta\boldsymbol{B}$ can be neglected from the linearised induction equation except where multiplied by the large value $\eta$ (or $Rm^{-1}$).
It follows that $\delta\boldsymbol{J} = \mu^{-1}\nabla \times \delta\boldsymbol{B} =-(\eta\mu)^{-1}(\boldsymbol{B\times\delta v})_{\rm solenoidal}$,  whence the perturbed Lorentz force $\delta\boldsymbol{J} \times \boldsymbol{B}$ becomes $(\eta\mu)^{-1}\boldsymbol{B\times}(\boldsymbol{B\times\delta v})_{\rm solenoidal}$.
(We are assuming that $\boldsymbol{J}=0$ in the base state.)
Hence the linearised flow does not depend on the background field strength $B$ and diffusivity $\eta$ independently, but only through combinations that scale as $B^2/\eta$.
If an inductionless instability exists, then it persists in the limit that $Pm,\, Rm\to0$ at fixed $Ha\ (\propto B/\sqrt{\eta\nu})$ and $\Lambda\ (\propto B^2/\eta\Omega)$.

Inductionless MHD instabilities are of doubtful relevance to thin astrophysical discs.
HMRI requires a larger-than-keplerian ratio of shear to rotation \cite{liu06b}.
More generally, an inductionless instability for which $Rm<1$ but $\Lambda\gtrsim 1$ will have Lehnert number $Le^2=B^2/\mu\rho(r\Omega)^2=\Lambda/Rm > 1$.
Yet in thin astrophysical discs, the magnetic energy per unit mass ($B^2/2\mu\rho$) is at most comparable to the thermal energy, which in turn is smaller than the orbital energy [$(r\Omega)^2/2$] by a factor $\sim (h/r)^2\ll 1$.
Even if $Rm$ is based on the thickness $h$ rather than $r$, i.e. $Rm\equiv h^2\Omega/\eta$, the magnetic pressure would still have to be larger than the thermal energy.

SMRI is certainly not inductionless: it requires $Rm>1$ at least.
One would like to see the recently reported detections of SMRI \cite{wang22a,wang22b} verified and studied more distinctly at higher $Rm$, perhaps in the upcoming DRESDYN experiment~\cite{stefani19,mishra22} with properly designed axial conducting boundaries despite its favorable large aspect ratio.


\section{Conclusion}

Over a long history of more than 130 years, quasi-keplerian Taylor-Couette flow received scant attention until prompted by astrophysical interest about two decades ago. 
Substantial progress has since been made. 
In hydrodynamics, the flow is extremely sensitive to axial boundaries. 
When proper care is taken to minimize Ekman effects, e.g. by using segmented, independently driven endrings, essentially laminar flow prevails at shear Reynolds numbers as large as $Re\approx 10^6$. 
This is remarkable, as few other shear flows are known to be so stable, and has already suggested applications~\cite{ji19}. 
It remains to be seen if stability persists to even higher $Re$, but the results already cast serious doubt on purely hydrodynamic accretion-disc turbulence at least in the unstratified and incompressible limit.

In magnetohydrodynamics, SSL instabilities have generated some confusion.
Careful endcap design has provided opportunities not only to increase SMRI amplitudes, however, but also to lower threshold magnetic Reynolds numbers $Rm=Pm Re$ via imperfect bifurcations.
Both developments have facilitated recently reported detections of SMRI, three decades after its astrophysical importance was first recognized. 
The experimental exploration of nonlinear MRI is a promising field that is only beginning.

Other astrophysically motivated Taylor-Couette work exists that is not covered by this mini-review. 
This includes stratorotational instability \cite[and references therein]{meletti21}, which may be relevant to accretion discs with vertical thermal gradients and attendant vertical shear. 
There is also work on TC experiments with plasma, where flow is driven by biasing electrodes across multipolar edge-mounted magnetic fields \cite{collins12}, or by radial voltages and currents across axial magnetic field~\cite{flanagan20}. 
A swirling partially ionized gas experiment in a cylindrical geometry using an injection-pumping system and spiral antennas mounted on endcaps to transmit RF power for plasma production can be used to study SMRI with Hall effects and ambipolar diffusion~\cite{ji11,secunda22}.
The tensorial versions of Ohms' Law in these plasma MRI experiments may be relevant to weakly ionised protoplanetary discs \cite{Lesur+2022}.

\enlargethispage{20pt}

\ethics{N/A.}

\dataccess{N/A.}

\aucontribute{Both authors discussed, drafted, and revised manuscript.}

\competing{No competing interests.}

\funding{Authors acknowledge support by U.S. Department of Energy via contract no. DE-AC0209CH11466 and National Science Foundation via grant no. AST-2108871.}

\ack{Authors acknowledge assistance by Dr. Yin Wang for reformatting Fig.6. HJ thanks the Isaac Newton Institute for Mathematical Sciences, Cambridge, for support during the programme DYT2 (via  EPSRC grant no EP/R014604/1) where some work on this paper was undertaken.}

\disclaimer{N/A.}


\bibliography{MRI}

\begin{thebibliography}{99}

\bibitem{mestel+spitzer56}
{Mestel} L, {Spitzer}, L. J. 1956  {Star formation in magnetic dust clouds}.
  {\em Mon. Not. R. Astron. Soc.} \textbf{116}, 503.

\bibitem{zhao+20}
{Zhao} B, {Tomida} K, {Hennebelle} P, {Tobin} JJ, {Maury} A, {Hirota} T,
  {S{\'a}nchez-Monge} {\'A}, {Kuiper} R, {Rosen} A, {Bhandare} A, {Padovani} M,
  {Lee} YN. 2020  {Formation and Evolution of Disks Around Young Stellar
  Objects}. {\em Space Science Reviews} \textbf{216}, 43.

\bibitem{andrews+2018}
{Andrews} SM, {Huang} J, {P{\'e}rez} LM, {Isella} A, {Dullemond} CP, {Kurtovic}
  NT, {Guzm{\'a}n} VV, {Carpenter} JM, {Wilner} DJ, {Zhang} S, {Zhu} Z,
  {Birnstiel} T, {Bai} XN, {Benisty} M, {Hughes} AM, {{\"O}berg} KI, {Ricci} L.
  2018  {The Disk Substructures at High Angular Resolution Project (DSHARP). I.
  Motivation, Sample, Calibration, and Overview}. {\em Astrophys. J. Lett.}
  \textbf{869}, L41.

\bibitem{frank2002accretion}
Frank J, King A, Raine D. 2002 {\em Accretion power in astrophysics}.
Cambridge university press.

\bibitem{zeldovich81}
Zeldovich Y. 1981  {On the Friction of Fluids Between Rotating Cylinders}. {\em
  Proc. Roy. Soc. London A} \textbf{374}, 299--312.

\bibitem{richard99}
Richard D, Zahn JP. 1999  {Turbulence in differentially rotating flows: What
  can be learned from the Couette-Taylor experiment.}. {\em Astron. Astrophys.}
  \textbf{347}, 734--738.

\bibitem{balbus91}
Balbus S, Hawley J. 1991  A powerful local shear instability in weakly
  magnetized disks. I - Linear analysis.. {\em Astrophys. J.} \textbf{376},
  214--222.

\bibitem{richard01}
Richard D. 2001 {\em {Instabilit\'es Hydrodynamiques dans les Ecoulements en
  Rotation Diff\'erentielle}}.
PhD thesis Universit\'e Paris 7.

\bibitem{ji01}
Ji H, Goodman J, Kageyama A. 2001  Magnetorotational instability in a rotating
  liquid metal annulus.. {\em Mon. Not. R. Astron. Soc.} \textbf{325}, L1--L5.

\bibitem{mallock1889}
Mallock A. 1889  Determination of the Viscosity of Water. {\em Proc. R. Soc.
  Lond. A} \textbf{45}, 126--132.

\bibitem{mallock1896}
Mallock A. 1896  Experiments on Fluid Viscosity. {\em Philos. Trans. R. Soc.}
  \textbf{187}, 41--56.

\bibitem{couette1890}
Couette M. 1890  {\'{E}tudes sur le frottement des liquides}. {\em Ann. Chim.
  Phys.} \textbf{6}, 433--510.

\bibitem{wendt33}
Wendt F. 1933  Turbulente {S}tr\"omungen zwischen zwei rotierenden konaxialen
  {Z}ylindern.. {\em Ing. Arch.} \textbf{4}, 577--595.

\bibitem{taylor36}
Taylor G. 1936  Fluid friction between rotating cylinders. I. Torque
  measurments.. {\em Proc. Roy. Soc. London A} \textbf{157}, 546--578.

\bibitem{deguchi17}
Deguchi K. 2017  Linear instability in Rayleigh-stable Taylor-Couette flow.
  {\em Physical Review E} \textbf{95}, 021102.

\bibitem{mamatsashvili19}
Mamatsashvili G, Stefani F, Hollerbach R, R\"udiger G. 2019  Two types of
  axisymmetric helical magnetorotational instability in rotating flows with
  positive shear. {\em Physical Review Fluids} \textbf{4}, 103905.

\bibitem{tillmark96}
Tillmark N, Alfredsson PH. 1996  Experiments on rotating plane {C}ouette flow.
  In Gavrilakis S, Machiels L, Monkewitz PA, editors, {\em Advances in
  Turbulence VI} pp. 391--394. Kluwer.

\bibitem{lesur05}
Lesur G, Longaretti PY. 2005  {On the relevance of Subcritical Hydrodynamic
  Turbulence to Accretion Disk Transport}. {\em Astron. Astrophys.}
  \textbf{444}, 25--44.

\bibitem{feldmann22}
Feldmann D, et~al.. 2022  Routes to turbulence in TC flow. {\em submitted to
  Philos. Trans. R. Soc. A}.

\bibitem{Belyaev+Rafikov2012}
{Belyaev} MA, {Rafikov} RR. 2012  {Supersonic Shear Instabilities in
  Astrophysical Boundary Layers}. {\em Astrophys. J.} \textbf{752}, 115.

\bibitem{Coleman+2022}
{Coleman} MSB, {Rafikov} RR, {Philippov} AA. 2022  {Boundary layers of
  accretion discs: Discovery of vortex-driven modes and other waves}. {\em Mon.
  Not. R. Astron. Soc.} \textbf{509}, 440--462.

\bibitem{rayleigh1916}
Rayleigh L. 1916  {On the dynamics of rotating fluid}. {\em Proc. R. Soc. Lond.
  A} \textbf{93}, 148--154.

\bibitem{taylor1923}
Taylor G. 1923  {Stability of a viscous liquid contained between two rotating
  cylinders}. {\em Philos. Trans. R. Soc. Lond. A} \textbf{223}, 289--343.

\bibitem{beckley02}
Beckley H. 2002 {\em {Measurements of Annular Couette Flow Stability at the
  Fluid Reynolds Number Re = $4.4 \times 10^6$: The Fluid Dynamic Precursor to
  a Liquid Sodium $\alpha \omega$ Dynamo}}.
PhD thesis New Mexico Institute of Mining and Technology.

\bibitem{kageyama04}
Kageyama A, Ji H, Goodman J, Chen F, Shoshan E. 2004  Numerical and
  Experimental Investigation of Circulation in Short Cylinders. {\em J. Phys.
  Soc. Jpn.} \textbf{73}, 2424--2437.

\bibitem{ji06}
Ji H, Burin M, Schartman E, Goodman J. 2006  Hydrodynamic turbulence cannot
  transport angular momentum effectively in astrophysical disks.. {\em Nature.}
  \textbf{444}, 343--346.

\bibitem{schartman12}
Schartman E, Ji H, Burin M, Goodman J. 2012  Stability of quasi-Keplerian shear
  flow in a laboratory experiment. {\em Astron. Astrophys.} \textbf{543}, A94.

\bibitem{obabko08}
{Obabko} AV, {Cattaneo} F, {F Fischer} P. 2008  {The influence of horizontal
  boundaries on Ekman circulation and angular momentum transport in a
  cylindrical annulus}. {\em Physica Scripta Vol. T} \textbf{132}, 014029.

\bibitem{paoletti11}
{Paoletti} MS, {Lathrop} DP. 2011  {Angular momentum transport in turbulent
  flow between independently rotating cylinders}. {\em Phys. Rev. Lett.}
  \textbf{106}, 024501.

\bibitem{paoletti12}
Paoletti M, van Gils D, Dubrulle B, Sun C, Lohse D, Lathrop D. 2012  Angular
  momentum transport and turbulence in laboratory models of Keplerian flows.
  {\em Astron. Astrophys.} \textbf{547}, A64.

\bibitem{avila12}
{Avila} M. 2012  {Stability and Angular-Momentum Transport of Fluid Flows
  between Corotating Cylinders}. {\em Phys. Rev. Lett.} \textbf{108}, 124501.

\bibitem{edlund14}
Edlund E, Ji H. 2014  Nonlinear stability of laboratory quasi-Keplerian flows.
  {\em Phys. Rev. E} \textbf{89}, 021004.

\bibitem{nordsiek15}
Nordsiek F, Huisman SG, van~der Veen RCA, Sun C, Lohse D, Lathrop DP. 2015
  {Azimuthal velocity profiles in Rayleigh-stable Taylor-Couette flow and
  implied axial angular momentum transport}. {\em Journal of Fluid Mechanics}
  \textbf{774}, 342--362.

\bibitem{edlund15}
Edlund E, Ji H. 2015  Reynolds number scaling of the influence of boundary
  layers on the global behavior of laboratory quasi-Keplerian flows. {\em Phys.
  Rev. E} \textbf{92}, 043005.

\bibitem{lopez17}
Lopez JM, Avila M. 2017  {Boundary-layer turbulence in experiments on
  quasi-Keplerian flows}. {\em Journal of Fluid Mechanics} \textbf{817},
  21--34.

\bibitem{balbus96}
Balbus SA, Hawley JF, Stone JM. 1996  Nonlinear stability, hydrodynamical
  turbulence, and transport in disks. {\em Astrophys. J.} \textbf{467}, 76--86.

\bibitem{hawley99}
Hawley J, Balbus S, Winters W. 1999  Local hydrodynamic stability of accretion
  disks.. {\em ApJ} \textbf{518}, 394.

\bibitem{coles65}
Coles D. 1965  Transition in circular Couette flow. {\em J. Fluid Mech}
  \textbf{21}, 385--425.

\bibitem{schartman09}
Schartman E, Ji H, Burin M. 2009  {Development of a Couette-Taylor flow device
  with active minimization of secondary circulation}. {\em Rev. Sci. Instrum.}
  \textbf{80}, 024501.

\bibitem{taylor36b}
Taylor G. 1936  Fluid Friction Between Rotating Cylinders. II - Distribution of
  Velocity Between Concentric when Outer One is Rotating and Inner One is at
  Rest. {\em Proc. R. Soc. Lond. A} \textbf{157}, 565--578.

\bibitem{hollerbach04}
{Hollerbach} R, {Fournier} A. 2004  {End-effects in rapidly rotating
  cylindrical Taylor-Couette flow}. In {\em AIP Conf. Proc. 733: MHD Couette
  Flows: Experiments and Models} pp. 114--121.

\bibitem{goodman02}
Goodman J, Ji H. 2002  Magnetorotational instability of dissipative Couette
  flow.. {\em J. Fluid Mech.} \textbf{462}, 365.

\bibitem{burin06}
Burin MJ, Schartman E, Ji H, Cutler R, Heitzenroeder P, Liu W, Morris L,
  Raftopolous S. 2006  Reduction of {E}kman Circulation within a Short Circular
  Couette Flow. {\em Experiments in Fluids} \textbf{40}, 962--966.

\bibitem{ostilla14}
Ostilla-M\'onico R, Verzicco R, Grossmann S, Lohse D. 2014  Turbulence decay
  towards the linearly stable regime of Taylor-Couette flow. {\em Journal of
  Fluid Mechanics} \textbf{748}, R3.

\bibitem{shi17}
Shi L, Hof B, Rampp M, Avila M. 2017  Hydrodynamic turbulence in
  quasi-Keplerian rotating flows. {\em Physics of Fluids} \textbf{29}.

\bibitem{Hayashi1981}
{Hayashi} C. 1981  {Structure of the Solar Nebula, Growth and Decay of Magnetic
  Fields and Effects of Magnetic and Turbulent Viscosities on the Nebula}. {\em
  Progress of Theoretical Physics Supplement} \textbf{70}, 35--53.

\bibitem{mullin2011experimental}
Mullin T. 2011  Experimental studies of transition to turbulence in a pipe.
  {\em Annual Review of Fluid Mechanics} \textbf{43}, 1--24.

\bibitem{Narayan+1987}
{Narayan} R, {Goldreich} P, {Goodman} J. 1987  {Physics of modes in a
  differentially rotating system - Analysis of the shearing sheet}. {\em Mon.
  Not. R. Astron. Soc.} \textbf{228}, 1--41.

\bibitem{Lesur+2022}
{Lesur} G, {Ercolano} B, {Flock} M, {Lin} MK, {Yang} CC, {Barranco} JA,
  {Benitez-Llambay} P, {Goodman} J, {Johansen} A, {Klahr} H, {Laibe} G, {Lyra}
  W, {Marcus} P, {Nelson} RP, {Squire} J, {Simon} JB, {Turner} N, {Umurhan} OM,
  {Youdin} AN. 2022  {Hydro-, Magnetohydro-, and Dust-Gas Dynamics of
  Protoplanetary Disks}. {\em arXiv e-prints} p. arXiv:2203.09821.

\bibitem{balbus92a}
{Balbus} SA, {Hawley} JF. 1992  {Is the Oort A-Value a Universal Growth Rate
  Limit for Accretion Disk Shear Instabilities?}. {\em Astrophys. J.}
  \textbf{392}, 662.

\bibitem{hung19}
Hung DMH, Blackman EG, Caspary KJ, Gilson EP, Ji H. 2019  Experimental
  confirmation of the standard magnetorotational instability mechanism with a
  spring-mass analogue. {\em Communications Physics} \textbf{2}, 7.

\bibitem{ruediger01}
{R{\"u}diger} G, {Zhang} Y. 2001  {MHD instability in differentially-rotating
  cylindric flows}. {\em Astron. Astrophys.} \textbf{378}, 302--308.

\bibitem{sisan04}
{Sisan} DR, {Mujica} N, {Tillotson} WA, {Huang} Y, {Dorland} W, {Hassam} AB,
  {Antonsen} TM, {Lathrop} DP. 2004  {Experimental Observation and
  Characterization of the Magnetorotational Instability}. {\em Phys. Rev.
  Lett.} \textbf{93}, 114502.

\bibitem{hollerbach05}
{Hollerbach} R, {R{\"u}diger} G. 2005  {New Type of Magnetorotational
  Instability in Cylindrical Taylor-Couette Flow}. {\em Phys. Rev. Lett.}
  \textbf{95}, 124501--+.

\bibitem{liu06a}
Liu W, Goodman J, Ji H. 2006  {Simulation of Magnetorotational Instability in a
  Magnetized Couette Flow}. {\em Astrophys. J.} \textbf{643}, 306.

\bibitem{liu08a}
Liu W. 2008  {Numerical study of the Magnetorotational Instability in the
  Princeton MRI Experiment}. {\em Astrophys. J.} \textbf{684}, 515.

\bibitem{liu06b}
Liu W, Goodman J, Herron I, Ji H. 2006  {Helical magnetorotational instability
  in magnetized Taylor-Couette flow}. {\em Phys. Rev. E} \textbf{74}, 056302.

\bibitem{liu07}
Liu W, Goodman J, Ji H. 2007  {Traveling waves in a magnetized Taylor-Couette
  flow}. {\em Phys. Rev. E} \textbf{76}, 016310.

\bibitem{stefani06}
Stefani F, Gundrum T, Gerbeth G, R{\"{u}}diger G, Schultz M, Szklarski J,
  Hollerbach R. 2006  Experimental Evidence for Magnetorotational Instability
  in a Taylor-Couette Flow under the Influence of a Helical Magnetic Field..
  {\em Phys. Rev. Lett.} \textbf{97}, 184502.

\bibitem{stefani09}
Stefani F, Gerbeth G, Gundrum T, Hollerbach R, Priede J, R{\"u}diger G,
  Szklarski J. 2009  Helical magnetorotational instability in a Taylor-Couette
  flow with strongly reduced Ekman pumping. {\em Physical Review E}
  \textbf{80}, 66303.

\bibitem{priede07}
Priede J, Grants I, Gerbeth G. 2007  Inductionless magnetorotational
  instability in a Taylor-Couette flow with a helical magnetic field. {\em
  Physical Review E - Statistical, Nonlinear, and Soft Matter Physics}
  \textbf{75}.

\bibitem{ruediger07b}
R\"{u}diger G, Hollerbach R, Gellert M, Schultz M. 2007  The azimuthal
  magnetorotational instability (AMRI). {\em Astronomische Nachrichten}
  \textbf{328}, 1158--1161.

\bibitem{hollerbach10}
Hollerbach R, Teeluck V, R\"udiger G. 2010  Nonaxisymmetric Magnetorotational
  Instabilities in Cylindrical Taylor-Couette Flow. {\em Phys. Rev. Lett.}
  \textbf{104}, 044502.

\bibitem{boldyrev09}
Boldyrev S, Huynh D, Pariev V. 2009  Analog of astrophysical magnetorotational
  instability in a Couette-Taylor flow of polymer fluids. {\em Physical Review
  E - Statistical, Nonlinear, and Soft Matter Physics} \textbf{80}.

\bibitem{bai15}
Bai Y, Crumeyrolle O, Mutabazi I. 2015  Viscoelastic Taylor-Couette instability
  as analog of the magnetorotational instability. {\em Physical Review E -
  Statistical, Nonlinear, and Soft Matter Physics} \textbf{92}.

\bibitem{bai21}
Bai Y, Vieu T, Crumeyrolle O, Mutabazi I. 2021  Viscoelastic
  Taylor‚ÄìCouette instability in the Keplerian regime. {\em Geophysical
  and Astrophysical Fluid Dynamics} \textbf{115}, 322--344.

\bibitem{nornberg10}
Nornberg M, Ji H, Schartman E, Roach A, Goodman J. 2010  Observation of
  Magnetocoriolis Waves in a Liquid Metal {T}aylor {C}ouette Experiment. {\em
  Phys. Rev. Lett.} \textbf{104}, 074501.

\bibitem{gissinger11}
Gissinger C, Ji H, Goodman J. 2011  {Instabilities in magnetized spherical
  Couette flow}. {\em Phys. Rev. E} \textbf{84}, 026308.

\bibitem{roach12}
{Roach} AH, {Spence} EJ, {Gissinger} C, {Edlund} EM, {Sloboda} P, {Goodman} J,
  {Ji} H. 2012  {Observation of a Free-Shercliff-Layer Instability in
  Cylindrical Geometry}. {\em Phys. Rev. Lett.} \textbf{108}, 154502.

\bibitem{roach13}
Roach A. 2013 {\em Velocity measurements and free-shear-layer instabilities in
  a rotating liquid metal}.
PhD thesis Princeton University.

\bibitem{spence12}
{Spence} EJ, {Roach} AH, {Edlund} EM, {Sloboda} P, {Ji} H. 2012  {Free
  magnetohydrodynamic shear layers in the presence of rotation and magnetic
  field}. {\em Phys. Plasmas} \textbf{19}, 056502.

\bibitem{gissinger12}
{Gissinger} C, {Goodman} J, {Ji} H. 2012  {The role of boundaries in the
  magnetorotational instability}. {\em Phys. Fluids} \textbf{24}, 074109.

\bibitem{seilmayer14}
Seilmayer M, Galindo V, Gerbeth G, Gundrum T, Stefani F, Gellert M, R{\"u}diger
  G, Schultz M, Hollerbach R. 2014  {Experimental Evidence for Nonaxisymmetric
  Magnetorotational Instability in a Rotating Liquid Metal Exposed to an
  Azimuthal Magnetic Field}. {\em Physical Review Letters} \textbf{113},
  024505.

\bibitem{wei16}
{Wei} X, {Ji} H, {Goodman} J, {Ebrahimi} F, {Gilson} E, {Jenko} F, {Lackner} K.
  2016  {Numerical simulations of the Princeton magnetorotational instability
  experiment with conducting axial boundaries}. {\em Phys. Rev. E} \textbf{94},
  063107.

\bibitem{winarto20}
Winarto HW, Ji H, Goodman J, Ebrahimi F, Gilson EP, Wang Y. 2020  Parameter
  space mapping of the Princeton magnetorotational instability experiment. {\em
  Physical Review E} \textbf{102}, 023113.

\bibitem{caspary18}
Caspary KJ, Choi D, Ebrahimi F, Gilson EP, Goodman J, Ji H. 2018  Effects of
  axial boundary conductivity on a free Stewartson-Shercliff layer. {\em Phys.
  Rev. E} \textbf{97}, 063110.

\bibitem{choi19}
Choi D, Ebrahimi F, Caspary KJ, Gilson EP, Goodman J, Ji H. 2019
  {Nonaxisymmetric simulations of the Princeton magnetorotational instability
  experiment with insulating and conducting axial boundaries}. {\em Physical
  Review E} \textbf{100}, 033116.

\bibitem{vernet21}
{Vernet} M, {Pereira} M, {Fauve} S, {Gissinger} C. 2021  {Turbulence in
  electromagnetically driven Keplerian flows}. {\em Journal of Fluid Mechanics}
  \textbf{924}, A29.

\bibitem{vernet22}
{Vernet} M, {Fauve} S, {Gissinger} C. 2022  {Angular Momentum Transport by
  Keplerian Turbulence in Liquid Metals}. {\em arXiv e-prints} p.
  arXiv:2206.14214.

\bibitem{wang22a}
Wang Y, Gilson E, Ebrahimi F, Goodman J, Ji H. 2022a  Observation of
  axisymmetric standard magnetorotational instability in the laboratory. {\em
  Phys. Rev. Lett.} \textbf{129}, 115001.

\bibitem{wang22b}
Wang Y, Gilson E, Ebrahimi F, Goodman J, Caspary K, Winarto H, Ji H. 2022b
  Identification of a non-axisymmetric mode in laboratory experiments searching
  for standard magnetorotational instability. {\em Nature Communications}
  \textbf{13}, 4679.

\bibitem{velikhov59}
Velikhov EP. 1959  Stability of an ideally conducting liquid flowing between
  cylinders rotating in a magnetic field.. {\em Sov. Phys. JETP} \textbf{36},
  995--998.

\bibitem{chandra60}
Chandrasekhar S. 1960  The stability of non-dissipative Couette flow in
  hydromagnetics.. {\em Proc. Nat. Acad. Sci.} \textbf{46}, 253--257.

\bibitem{balbus98}
Balbus S, Hawley J. 1998  Instability, turbulence, and enhanced transport in
  accretion disks. {\em Rev. Mod. Phys.} \textbf{70}, 1--53.

\bibitem{pessah08}
{Pessah} ME, {Chan} Ck. 2008  {Viscous, Resistive Magnetorotational Modes}.
  {\em Astrophys. J.} \textbf{684}, 498--514.

\bibitem{donnelly60}
Donnelly RJ, Ozima M. 1960  Hydromagnetic stability of flow between rotating
  cylinders.. {\em Phys. Rev. Lett.} \textbf{4}, 497--498.

\bibitem{willis02}
{Willis} AP, {Barenghi} CF. 2002  {Magnetic instability in a sheared azimuthal
  flow}. {\em Astron. Astrophys.} \textbf{388}, 688--691.

\bibitem{noguchi02}
{Noguchi} K, {Pariev} VI, {Colgate} SA, {Beckley} HF, {Nordhaus} J. 2002
  {Magnetorotational Instability in Liquid Metal Couette Flow}. {\em Astrophys.
  J.} \textbf{575}, 1151--1162.

\bibitem{knobloch05}
{Knobloch} E, {Julien} K. 2005  {Saturation of the magnetorotational
  instability}. {\em Physics of Fluids} \textbf{17}, 094106--1.

\bibitem{Umurhan+2007}
{Umurhan} OM, {Regev} O, {Menou} K. 2007  {Nonlinear saturation of the
  magnetorotational instability near threshold in a thin-gap Taylor-Couette
  setup}. {\em Phys. Rev. E} \textbf{76}, 036310.

\bibitem{willis02b}
Willis AP, Barenghi CF. 2002  A Taylor-Couette dynamo. {\em Astronomy and
  Astrophysics} \textbf{393}, 339--343.

\bibitem{clark+oishi17}
{Clark} SE, {Oishi} JS. 2017  {The Weakly Nonlinear Magnetorotational
  Instability in a Global, Cylindrical Taylor-Couette Flow}. {\em Astrophys.
  J.} \textbf{841}, 2.

\bibitem{stewartson57}
{Stewartson} K. 1957  {On almost rigid rotations}. {\em J. Fluid Mech.}
  \textbf{3}, 17--26.

\bibitem{shercliff53}
{Shercliff} JA. 1953  Steady motion of conducting fluids in pipes under
  transverse magnetic fields. {\em Math. Proc. Cam. Phi. Soc.} \textbf{49},
  136.

\bibitem{lehnert55}
Lehnert B. 1955  An instability of laminar flow of mercury caused by an
  external magnetic field. {\em Proc. R. Soc. A.} \textbf{233}, 299--302.

\bibitem{hollerbach01}
Hollerbach R, Skinner S. 2001  Instabilities of magnetically induced shear
  layers and jets.. {\em Proc. R. Soc. Lond. A} \textbf{457}, 785--802.

\bibitem{knobloch96}
Knobloch E. 1996  Symmetry and instability in rotating hydrodynamic and
  magnetohydrodynamic flows. {\em Phys. Fluids} \textbf{8}, 1446--1454.

\bibitem{oishi20}
{Oishi} JS, {Vasil} GM, {Baxter} M, {Swan} A, {Burns} KJ, {Lecoanet} D, {Brown}
  BP. 2020  {The magnetorotational instability prefers three dimensions}. {\em
  Proceedings of the Royal Society of London Series A} \textbf{476}, 20190622.

\bibitem{ruediger07}
{R{\"u}diger} G, {Hollerbach} R. 2007  Comment on``Helical magnetorotational
  instability in magnetized Taylor-Couette flow''. {\em Phys. Rev. E}
  \textbf{76}, 068301.

\bibitem{velikhov06b}
{Velikhov} EP, {Ivanov} AA, {Lakhin} VP, {Serebrennikov} KS. 2006
  {Magneto-rotational instability in differentially rotating liquid metals}.
  {\em Physics Letters A} \textbf{356}, 357--365.

\bibitem{stefani19}
Stefani F, Gailitis A, Gerbeth G, Giesecke A, Gundrum T, R{\"u}diger G,
  Seilmayer M, Vogt T. 2019  The DRESDYN project: liquid metal experiments on
  dynamo action and magnetorotational instability. {\em Geophysical and
  Astrophysical Fluid Dynamics} \textbf{113}, 51--70.

\bibitem{mishra22}
Mishra A, Mamatsashvili G, Stefani F. 2022  Nonlinear Simulations of
  Magnetorotational Instability: Scaling Properties and Relation to Upcoming
  DRESDYN-MRI Experiment. submitted.

\bibitem{ji19}
Ji H, Cohen A, Efthimion P, Edlund E, Gilson E. 2019  Advanced liquid
  centrifuge using differentially rotating cylinders and optimized boundary
  conditions, and methods for the separation of fluids. {\em U.S. Patent Number
  10,300,410}.

\bibitem{meletti21}
Meletti G, Abide S, Viazzo S, Krebs A, Harlander U. 2021  Experiments and
  long-term high-performance computations on amplitude modulations of
  strato-rotational flows. {\em Geophysical and Astrophysical Fluid Dynamics}
  \textbf{115}, 297--321.

\bibitem{collins12}
{Collins} C, {Katz} N, {Wallace} J, {Jara-Almonte} J, {Reese} I, {Zweibel} E,
  {Forest} CB. 2012  {Stirring Unmagnetized Plasma}. {\em Phys. Rev. Lett.}
  \textbf{108}, 115001.

\bibitem{flanagan20}
{Flanagan} K, {Milhone} J, {Egedal} J, {Endrizzi} D, {Olson} J, {Peterson} EE,
  {Sassella} R, {Forest} CB. 2020  {Weakly Magnetized, Hall Dominated Plasma
  Couette Flow}. {\em Phys. Rev. Lett.} \textbf{125}, 135001.

\bibitem{ji11}
{Ji} H. 2011  {Current status and future prospects for laboratory study of
  angular momentum transport relevant to astrophysical disks}. In {Bonanno} A,
  {de Gouveia Dal Pino} E, {Kosovichev} AG, editors, {\em IAU Symposium} vol.
  274{\em IAU Symposium} pp. 18--25.

\bibitem{secunda22}
Secunda A, Donnel P, Ji H, Gift D, Goodman J. 2022  Magnetorotational
  Instability in a Swirling Partially Ionized Gas. {\em to be submitted}.

\end{thebibliography}
\bibliographystyle{RS}
\end{document}